\documentclass[aps,prb,reprint,groupedaddress,superscriptaddress,longbibliography]{revtex4-2}
\usepackage{amsmath}
\usepackage{amssymb}
\usepackage{graphicx}
\usepackage{subfigure}
\usepackage{comment}
\usepackage{ntheorem}
\usepackage{tabularx}
\usepackage{color}
\newcommand{\wsx}{\color {black}}
\newcommand{\zb}{\color {black}}

\begin{document}
\newtheorem{Definition}{Definition}[subsection]
   \title{  General theory for infernal points in non-Hermitian systems}
   \author{Shu-Xuan Wang}
   \email{wangshx65@mail.sysu.edu.cn}
   \affiliation{Guangdong Provincial Key Laboratory of Magnetoelectric Physics and Devices, School of Physics, Sun Yat-sen University, Guangzhou 510275, China}
   \author{Zhongbo Yan}
   \email{yanzhb5@mail.sysu.edu.cn}
   \affiliation{Guangdong Provincial Key Laboratory of Magnetoelectric Physics and Devices, School of Physics, Sun Yat-sen University, Guangzhou 510275, China}

   \date{\today}

   \begin{abstract}
     The coalescence of eigenstates is a unique phenomena in non-Hermitian systems. Remarkably, it has been
     noticed in some non-Hermitian systems under open boundary conditions that the whole set of eigenstates can
     coalesce to only a few  eigenstates. In the parameter space, the point at which
     such a  coalescence of macroscopic eigenstates occurs is dubbed as an infernal point.
     In this paper, based on the non-Bloch band theory and amoeba formulation, we establish
     the criteria for the presence of infernal points in one-dimensional and higher dimensional
     open-boundary non-Hermitian systems. {\zb In addition}, we find an explanation of the extreme localization of the wave functions
     and unveil the mechanism for the coalescence of enormous eigenstates at the infernal points.
     Our work provides a general theory for infernal points in open-boundary non-Hermitian systems in arbitrary dimensions,
     and hence paves the way to study the intriguing infernal points systematically.
   \end{abstract}

   \maketitle

   \emph{Introduction}---Non-Hermitian systems have garnered tremendous attention in various research fields over the past decades\cite{Heiss2012EPreview,Ozdemir2019EPreview,Alu2019EP,Ding2022EPreview,Ganainy2018review,39,Ashida2020review,Bergholtz2021review,
   Okuma2023review,Huang2024review,Halder_2023,PhysRevB.109.115407,manna2023inner}, and a series of non-Hermitian systems exhibiting intriguing phenomena have been implemented in
   optical systems\cite{Peng2014PT,40,Zhen2015EPring,Wang2021EP}, ultracold atomic systems\cite{41,Ren2022}, and acoustic systems\cite{Zhu2018EP,42,43,Zhang2021winding,Zhang2021NHSE,Zhou2023geometry}.
   Despite the fact that the hermiticity of observable operators is a basic assumption in quantum mechanics,
   the non-Hermitian Hamiltonian provides an effective description for open systems involving gain and loss\cite{1,2,moiseyev2011non}, and it is discovered that non-hermiticity can give rise to diverse fascinating phenomena without counterparts in Hermitian systems\cite{18,25,26,27,28,Liu2019NHSOTI,29,33,34,35,36,37,44,45,48,49,50,51,52,53,54,55,56,57,58,62}. Notable examples
   include the coalescence of eigenstates\cite{Heiss1991,Heiss1999,Demange2012EP},
   non-Hermitian skin effect\cite{Yao2018GBZ,Yao2018Chern,5,6,7,8,9,10,11,46,47,59,60,61,PhysRevB.104.165117,PhysRevB.107.155430,PhysRevA.109.L061501}, non-Bloch $\mathcal{PT}$ symmetry breaking\cite{12,13,14}, and edge burst\cite{15,16,17}, to name a few.  Among them, the coalescence of eigenstates is a rather intriguing phenomenon
   specific to non-Hamiltonian systems. As is known, an Hermitian Hamiltonian is always
   diagonalizable, and the number of orthogonal eigenstates is equal to the
   {\zb dimension of the Hamiltonian}. In contrast, a non-Hermitian Hamiltonian in some cases
   can only be diagonalized to the Jordan canonical form. Under such circumstances,
   some eigenstates will coalesce to one eigenstate, rendering the dimension of the Hilbert space spanned
   by the eigenstates smaller than the dimension of the Hamiltonian. In a periodic system,
   the coalescence of eiegenstates is associated with the degeneracies of complex
   energy bands, which are known as exception points (EPs){\zb\cite{Heiss2012EPreview,
   Ozdemir2019EPreview,Alu2019EP,Ding2022EPreview}}. As a band degeneracy
   in general only involves a limited number of bands, the defectiveness of eigenstates
   at one EP is finite{\zb\cite{Ding2016EP,Xu2017EPring,Yang2019EPlink,Johan2018link,Johan2019EPknot,
   Kawabata2019EPclassification,Zhang2020HOEP,Hu2021EPknot,Mandal2021HOEP,Delplace2021HOEP,Liu2021EPring}}.

   In non-Hermitian systems, it is known that, owing to the skin effect\cite{Yao2018GBZ},
   the energy spectra under open boundary conditions (OBCs)
   can be drastically different from those under periodic boundary conditions (PBCs).  Remarkably,
   it has been noticed in some open-boundary non-Hermitian systems that an extensive number of eigenstates can
   coalesce to only a few eigenstates at some points in the parameter space\cite{Martinez2018IP,Kunst2019IP,Longhi2020IP,Terrier2020IP,31,32}. In other words,
   the defectiveness of eigenstates is macroscopic and increases with the system size. Such exotic points
   are dubbed as infernal points (IPs)\cite{31,Denner_2023}. Owing to the much more serious defectiveness, a system
   near an IP is expected to respond to external fields even more sensitively than a system near an
   EP{\zb\cite{Wiersig2014EP,Chen2017EP,Hodaei2017HOEP}}, a property
   holding promise for applications in diverse fields\cite{Hokmabadi2019EP,Lai2019EP,Kononchuk2022EP}.
   Thus far, the IPs have only been theoretically studied in
   a few simple models, and the method applied to determine the IPs is
   to calculate the characteristic polynomials and solve the system under OBCs directly\cite{Martinez2018IP,Kunst2019IP,Longhi2020IP,Terrier2020IP,31,32}.
   However, this approach is not efficient and becomes challenging for non-Hermitian
   systems with dimension higher than one. The lack of a general theory has greatly impeded
   the progress in understanding the IPs to a level comparable to the EPs.

   \par

   In this paper, based on the one-dimensional (1D) Hatano-Nelson (HN) model\cite{Hatano1996}, we find that the presence of IPs is simply connected
   to the contraction (or expansion) of the generalized Brillouin zone~(GBZ) to the origin (infinity).
   With the picture in mind, we build up an understanding of the extreme localization of wave functions
   and the mechanism of the coalescence of enormous eigenstates at the IPs. In terms of the non-Bloch band theory (NBBT)\cite{33,34,35,36,Yao2018GBZ},
   we further obtain the criteria for the presence of IPs analytically, and then {\zb exemplify the validity of} the theory
   in the non-Hermitian Su-Schrieffer-Heeger (SSH) model. At the end, a generalization of the criteria to higher dimensions is provided through the
   amoeba formulation\cite{37,YANG20221865,pnas.2302572120,yang2024anatomyhigherordernonhermitianskin,HU2024}.
   \par

   \emph{IPs in 1D non-Hermitian systems}---We start with the HN model\cite{Hatano1996},
     \begin{equation}
          H_{HN} = \sum_{n} t_R c_{n+1}^{\dagger} c_n + t_L c_{n-1}^{\dagger} c_n,
          \label{1}
     \end{equation}
    where $n \in [1,L]$ is the position of the site, $c_{n}$~($c_{n}^{\dagger}$) is the annihilation~(creation) operator at the
    $n$th site, and $t_R$ and $t_L$ are hopping amplitudes. Under OBCs and choosing the natural basis $\psi=(c_{1},...,c_{L})^{T}$,
    the Hamiltonian can be rewritten as $H_{HN}=\psi^{\dag}\mathcal{H}_{HN}\psi$, with
      \begin{equation}
        \mathcal{H}_{HN} =
          \begin{pmatrix}
              0  &  t_L &    &   &       \\
              t_R  &  0   & t_L &   &    \\
                   &   \ddots & \ddots  & \ddots &  \\
                   &   &  t_R  &   0  & t_L   \\
                   &   &       &   t_R  &  0
          \end{pmatrix}_{L \times L}.
          \label{2}
      \end{equation}
   Solving the Schr\"{o}dinger equation $\mathcal{H}_{HN} \Psi = E \Psi$, we
   find that if $t_R = 0$ or $t_L = 0$, there is only one eigenstate and all eigenergies coalesce
   at $E=0$, suggesting that an IP appears at these two limits. To be specific,
   when $t_R = 0$, we find that the only one eigenstate is of the form $\Psi_L = (1,0,0,\cdots,0)^T$;
   On the other hand, when $t_L = 0$, the only one eigenstate is given by $\Psi_R = (0,0,0,\cdots,1)^T$.
   A salient property of the wave functions at the IPs is the extreme localization at the left or right boundary.

   \begin{figure*}
        \centering
        \subfigure[]{\includegraphics[scale=0.41]{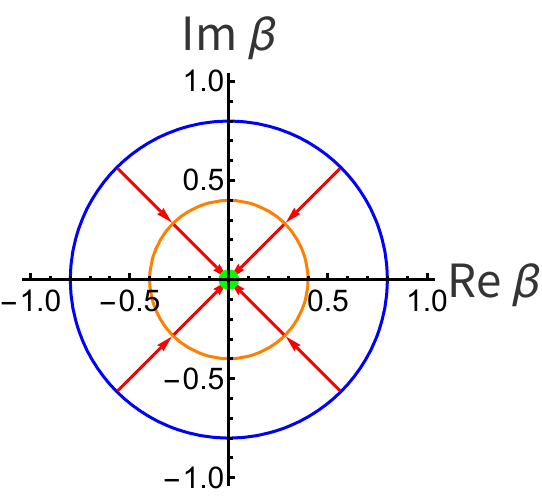} \label{fig1a}}
        \quad
        \subfigure[]{\includegraphics[scale=0.43]{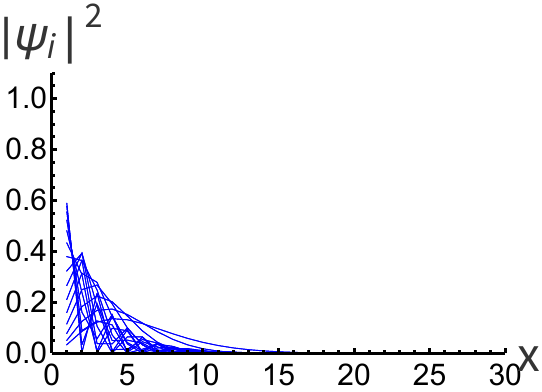} \label{figlb}}
        \quad
        \subfigure[]{\includegraphics[scale=0.43]{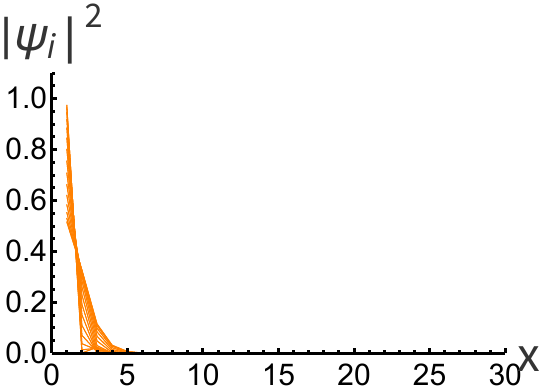} \label{fig1c}}
        \quad
        \subfigure[]{\includegraphics[scale=0.43]{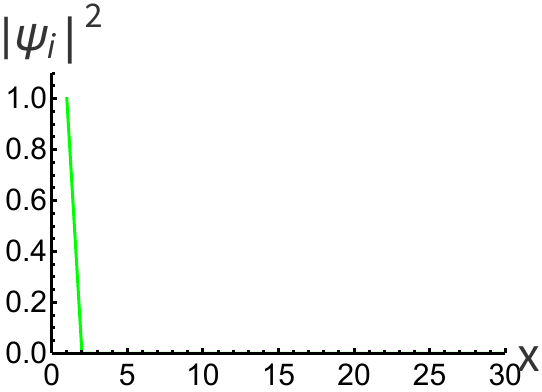} \label{fig1d}}
        \quad
        \subfigure[]{\includegraphics[scale=0.41]{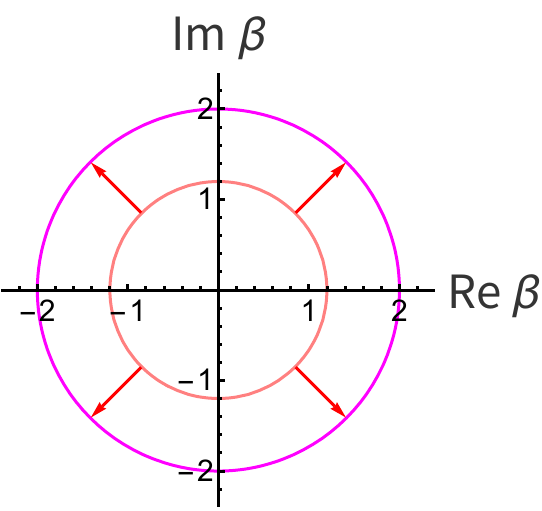} \label{fig1e}}
        \quad
        \subfigure[]{\includegraphics[scale=0.43]{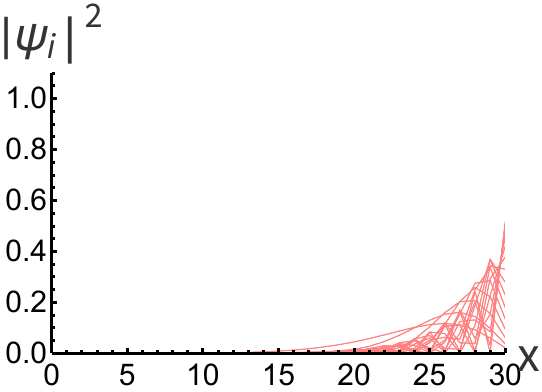} \label{figlf}}
        \quad
        \subfigure[]{\includegraphics[scale=0.43]{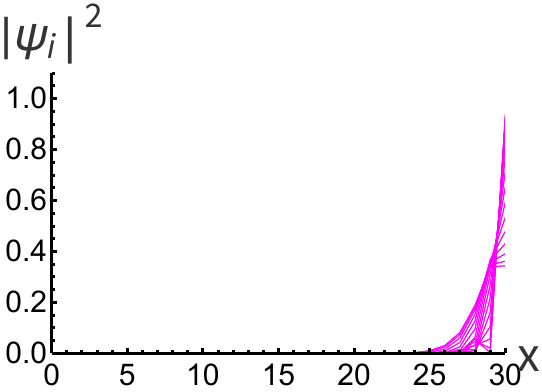} \label{fig1g}}
        \quad
        \subfigure[]{\includegraphics[scale=0.43]{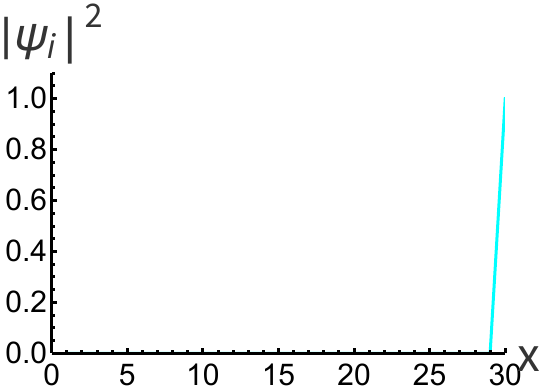} \label{fig1h}}
        \caption{GBZs and density amplitudes of wave functions for the HN model with the length $L = 30$ under different parameters. In (a), (b), (c) and (d), we set $t_L =1$, and in (e), (f), (g) and (h), we set $t_R =1$.  In (a), The blue ring, orange ring and the greeen point are GBZs for $t_R = \frac{16}{25}$, $t_R = \frac{4}{25}$ and $t_R = 0$, respectively. (b), (c) and (d) are distribution of density amplitudes for wave {\zb functions} corresponding to the cases $t_R = \frac{16}{25}$, $t_R = \frac{4}{25}$ and $t_R = 0$, respectively. When $t_R$ varies from $\frac{16}{25}$ to $0$, the GBZ shrinks from a circle to a point, and {\zb all wave functions coalesce to one which is localized at $x=1$}. In (e), the pink ring is the GBZ of {\zb the }HN model for $t_L = \frac{25}{36}$ and the magenta ring is the GBZ for $t_L = \frac{1}{4}$. (f), (g) {\zb and} (h) are distribution of density amplitudes for wave {\zb functions} {\zb referring} to $t_L = \frac{25}{36}$, $t_L = \frac{1}{4}$ and $t_L = 0$, respectively. When $t_L$ varies from $\frac{25}{36}$ to $0$, the radius of the GBZ changes from $\frac{6}{5}$ to {\zb $\infty$}, and {\zb all wave functions coalesce to one which is localized at $x=L$}. }
        \label{fig1}
      \end{figure*}

   According to the NBBT\cite{Yao2018GBZ,33}, the GBZ of this system is a circle whose radius is $|\beta| = \sqrt{t_R/t_L}$.
   At the IP corresponding to $t_R = 0$ ($t_{L}=0$), $|\beta| = 0$ $(+\infty)$. The result from this toy model
   suggests that the presence of IPs is connected to the contraction (expansion) of
   the GBZ to the origin (infinity) of the complex plane, see an illustration in Fig.\ref{fig1}.
   The collapse of GBZ and non-Bloch bands at the IPs has also been found case by case in some
   other 1D two-band non-Hermitian systems\cite{Martinez2018IP,Kunst2019IP,Longhi2020IP,32}. Therefore, a natural question to ask is
   whether the presence of IPs is always accompanied with the contraction-to-origin or expansion-to-infinity of the GBZ?
   In this work, we address this question in affirmative. As the contraction-to-origin case and the expansion-to-infinity case
   are due to each other and only different in the boundary at which the wave functions are localized, below we will mainly focus on the former case
   to avoid repetition. Our general theory will use the following fact from the NBBT. Namely, the GBZ is a curve or a group of curves in the complex plane, and the origin always lies inside it or them\cite{36,53}. This fact implies that if the GBZ shrinks to a point,
   the point must be the origin of the complex plane.

   Now we start the general analysis. Let us consider a 1D non-Hermitian system described by the following general Hamiltonian,
      \begin{equation}
        H = \sum_{n=1}^{L} \sum_{l=-p}^{q} \mathbf{c}_{n+l}^{\dagger} h_l \mathbf{c}_{n},
        \label{7}
      \end{equation}
    where $L$ is the length of the system, $l$ characterizes the hopping range, and $h_l$ is a $r \times r$ matrix with
    $r$ counting the internal degrees of freedom, i.e., $\mathbf{c}_{n} = (c_{n,1}, c_{n,2}, \cdots, c_{n,r})^T$ with
    $c_{n,i}$ the annihilation operator for one of the degrees of freedom at the $n$th unit cell. Under OBCs and also
    choosing the natural basis, the matrix form of the Hamiltonian is
      \begin{equation}
          \mathcal{H} =
            \begin{pmatrix}
              h_0  &  h_{-1} & \cdots   & h_{-p}  &      &        \\
              h_1  &  h_0   & \ddots & \ddots  &  \ddots   &      \\
              \vdots   &   \ddots & \ddots  & \ddots & \ddots & h_{-p}     \\
               h_q    & \ddots  &  \ddots  &   \ddots  & \ddots   &  \vdots    \\
                   & \ddots  &   \ddots    &  \ddots  &  h_0   &  h_{-1}    \\
                   &         &    h_q      &  \cdots   &  h_1    &   h_0
            \end{pmatrix}_{rL \times rL} .
            \label{9}
      \end{equation}
    The general form of the wave function can be expressed as $\Phi = (\varphi_1^T, \varphi_2^T, \cdots, \varphi_L^T)^T$, where
    $\varphi_n^T =(\varphi_{n,1}, \varphi_{n,2}, \cdots, \varphi_{n,r})^T$
    denotes the components of the wave function at the $n$th unit cell. According to the NBBT, the wave function for the eigenstate with energy $E(\beta)$ in the deep bulk has the form\cite{36}
      \begin{equation}
        \varphi_n (E (\beta)) \simeq \beta^n \varphi (\beta),\,\beta\in\text{GBZ}.
        \label{12}
      \end{equation}
     This formula shows that if the GBZ shrinks to the origin of the complex plane, i.e., $|\beta|=0$,
     all components of the wave functions deep in the bulk vanish. This fact implies that there exists a real-space truncation for all the wave functions. In other words, $\exists s \in \mathbb{Z}$ and $ s > 0$,  such that for $\forall L \geqslant n>s$ and $j=\{1,2,\cdots,r\}$, $\varphi_{n,j} = 0$. A point to note is that the value of $s$ is independent of $L$.
    When the hopping range is only up to the nearest neighbor, i.e., $p=q=1$,
    we note that such a truncation is consistent with the fixed point of the recursion of the $\Lambda$ matrix given in Ref.\cite{32}.
    Therefore, the position of the truncation can be determined by calculating the fixed point (see more details in the Supplemental Material\cite{supplemental}).

    Because of the existence of a truncation, all eigenstates and the corresponding eigenenergies
    can be {\zb effectively} determined by the truncated Hamiltonian
      \begin{equation}
        \mathcal{H}_{T} =
          \begin{pmatrix}
            h_0  &  h_{-1} & \cdots   & h_{-p}  &      &        \\
            h_1  &  h_0   & \ddots & \ddots  &  \ddots   &      \\
            \vdots   &   \ddots & \ddots  & \ddots & \ddots & h_{-p}     \\
             h_q    & \ddots  &  \ddots  &   \ddots  & \ddots   &  \vdots    \\
                 & \ddots  &   \ddots    &  \ddots  &  h_0   &  h_{-1}    \\
                 &         &    h_q      &  \cdots   &  h_1    &   h_0
          \end{pmatrix}_{rs \times rs} .
          \label{14}
      \end{equation}
    As aforementioned, $s$ is finite and independent of $L$. This means that
    the system can have $rs$ linearly independent eigenstates at most even if $L \rightarrow \infty$, and
    all eigenstates are localized at finite sites near the boundary. Because the defectiveness of
    the eigenstates becomes infinity as $L \rightarrow \infty$, the system is at an IP.  If the GBZ expands to infinity,
    the phenomenon is similar. The only difference is that the truncation is at the other side of the system, as exemplified by
    the HN model.

    Now let us determine the condition for the contraction (expansion) of the GBZ to origin (infinity).
    For the general Hamiltonian given in Eq.\eqref{7}, the corresponding non-Bloch form is
      \begin{equation}
        \mathcal{H} (\beta) = \sum_{l=-p}^{q} h_l \beta^{-l}.
        \label{16}
      \end{equation}
    The characteristic polynomial of $\mathcal{H} (\beta)$ is
      \begin{equation}
        \begin{split}
          f(\beta,E) = \det \left[ \mathcal{H} (\beta) - E\mathbb{I}_{r \times r} \right]
                     = \sum_{m=-rq}^{rp} \alpha_{m} \beta^m,
        \end{split}
        \label{17}
      \end{equation}
    where $\alpha_m $ are expansion coefficients whose values depend on hopping amplitudes and $E$, and
    $\mathbb{I}_{r \times r}$ is the $r \times r$ identity matrix.
    Our finding is that, if the coefficients $\alpha_m$  satisfy
      \begin{equation}
        \alpha_{-rq} = \alpha_{-rq+1} = \cdots = \alpha_{-1} = 0,
        \label{18}
      \end{equation}
    then the GBZ shrinks to the origin. On the other hand, if $\alpha_m$ satisfy
      \begin{equation}
        \alpha_{1} = \alpha_{2} = \cdots = \alpha_{rp} = 0,
        \label{19}
      \end{equation}
    then the  GBZ expands to the infinity\cite{supplemental}.
    Therefore, the criterion for the presence of IPs for a generic 1D non-Hermitian system
    is that the coefficients of its characteristic polynomial satisfy Eq.\eqref{18} or Eq.\eqref{19}. Before
    ending this part, it is worth mentioning that the presence of TRS$^{\dag}$ (the symplectic-class non-Hermitian
    Hamiltonians) could connect
    $\alpha_{m}$ and $\alpha_{-m}$, and then the contraction to origin and the expansion
    to infinity can simultaneously happen for two TRS$^{\dag}$-related GBZs. In this case,
    the criterion for the presence of IPs has a small modification\cite{supplemental}.

\begin{figure}
        \centering
        \subfigure[]{\includegraphics[scale=0.42]{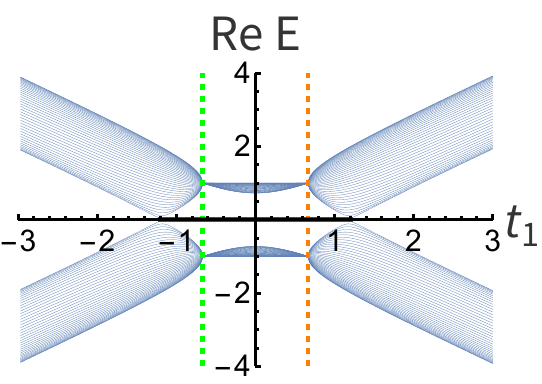} \label{fig2a}}
        \quad
        \subfigure[]{\includegraphics[scale=0.45]{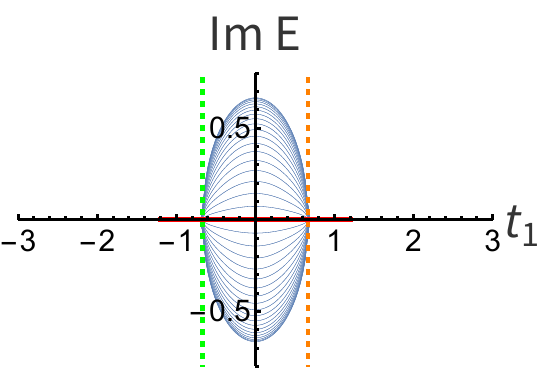} \label{fig2b}}
        \caption{Energy spectrum of the non-Hermitian SSH model under OBCs with $t_2 = 1$, $\gamma = \frac{2}{3}$ and $L = 40 $. (a) is the real part of the spectrum and (b) is the imaginary part. The vertical green (orange) dashed lines in both (a) and (b) correspond to $t_1 = -\frac{2}{3}$ ($\frac{2}{3}$),  and the horizontal red lines in both (a) and (b) correspond to the real and imaginary parts of the energies of the topological zero modes. At $t_1 = \pm \gamma$, this system has only $3$ energy levels, with $E = 0, \pm 1$.}
        \label{fig2}
      \end{figure}

    \emph{An example for illustration}---We use the non-Hermitian SSH model to exemplify
    the validity of the criteria given in Eq.\eqref{18} and Eq.\eqref{19}.  The explicit form of the Hamiltonian reads
      \begin{equation}
         \begin{split}
          H_{SSH} = \sum_{n=1}^{L}& (t_1 + \gamma) c_{n,A}^{\dagger} c_{n,B} + (t_1 - \gamma) c_{n,B}^{\dagger} c_{n,A}  \\
                    &+ t_2 (  c_{n+1,A}^{\dagger} c_{n,B} + c_{n-1,B}^{\dagger} c_{n,A}  ),
         \end{split}
         \label{20}
      \end{equation}
    where $n$ denotes the site position, $A$ and $B$ are labels of two sublattices,
    and $t_1$, $\gamma$ and $t_2$ are hopping parameters. It has been found in previous works that the spectrum of this system under OBCs will degenerate to $3$ discrete energy levels at $t_1 = \pm \gamma$\cite{Martinez2018IP}, as shown in Fig.\ref{fig2}.

    Solving the system directly, one finds that this system has an IP at $t_1 = \gamma$ or $-\gamma$\cite{32}. At $t_1 = \gamma$, the system has only three eigenstates whose explicit forms read
      \begin{equation}
          \begin{split}
            \Phi_{L,E=0} &= \left(1,0,0, 0,\cdots,0  \right)^T,
            \\
            \Phi_{L,E=t_2} &= \left(2 \gamma/t_2, 1,1,0, \cdots, 0  \right)^T,
            \\
            \Phi_{L,E=-t_2} &= \left(2 \gamma/t_2, -1,1,0, \cdots, 0  \right)^T.
          \end{split}
          \label{21}
      \end{equation}
    Similarly, for the IP at $t_1 = - \gamma$, the system also has only three eigenstates whose explicit forms read
      \begin{equation}
        \begin{split}
          \Phi_{R,E=0} &= \left(0,\cdots,0,0,1  \right)^T,
          \\
          \Phi_{R,E=t_2} =& \left(0 \cdots, 0, t_2/(2 \gamma) , t_2/(2 \gamma) ,1 \right)^T,
          \\
          \Phi_{R,E=-t_2} =& \left(0 \cdots, 0, t_2/(2 \gamma) , -t_2/(2 \gamma) ,1 \right)^T,
        \end{split}
        \label{22}
      \end{equation}
     According to Ref.\cite{Yao2018GBZ}, the GBZ of this Hamiltonian is a circle whose radius is given by
      \begin{equation}
          |\beta| = \sqrt{\frac{t_1 - \gamma}{t_1 + \gamma}}.
          \label{23}
      \end{equation}
    Apparently, $|\beta| = 0$ for $t_1 = \gamma$ and $|\beta| = \infty$ for $t_1 = - \gamma$.

    Now let us determine the characteristic polynomial.  The non-Bloch Hamiltonian of this system is given by
      \begin{equation}
        \mathcal{H}_{SSH} (\beta) =
           \begin{pmatrix}
              0     &     t_1 + \gamma + t_2 \beta^{-1}   \\
              t_1 - \gamma + t_2 \beta   &   0
           \end{pmatrix},
           \label{24}
      \end{equation}
    and the characteristic polynomial of $\mathcal{H}_{SSH} (\beta)$ is
      \begin{equation}
        \begin{split}
          f (\beta,E) =& \det \left[ \mathcal{H}_{SSH} (\beta) - E \mathbb{I}_{2 \times2} \right]
          \\
           =& -(t_1 t_2 + t_2 \gamma) \beta + (t_2 \gamma - t_1 t_2) \beta^{-1}
          \\
          & + (E^2 - t_1^2 -t_2^2 + \gamma^2).
        \end{split}
        \label{25}
      \end{equation}
    The coefficients are therefore given by
      \begin{equation}
         \begin{split}
          \alpha_{-2} &= \alpha_{2} \equiv 0,\quad
          \alpha_{-1} = t_2 \gamma - t_1 t_2,
          \\
          \alpha_{0} &= E^2 - t_1^2 -t_2^2 + \gamma^2,\quad
          \alpha_{1}= -(t_1 t_2 + t_2 \gamma).
         \end{split}
         \label{26}
      \end{equation}
    For the IP at $t_1 = \gamma$, one obtains $\alpha_{-2} = \alpha_{-1} = 0$, and for the IP at $t_1 = -\gamma$, one obtains $\alpha_{1} = \alpha_{2}= 0$. Apparently, the presence of IPs agrees with the criteria given in Eq.\eqref{18} and Eq.\eqref{19}.

    It is worth emphasizing that the GBZs for more general cases may just be a closed curve rather than a circle. In such cases, our theory is still valid for determining whether the system is at an IP or not. In the Supplemental Material, we provide a concrete example\cite{supplemental}.
    \par

    \emph{Generalized to higher dimensional non-Hermitian systems}---For higher dimensional non-Hermitian systems under OBCs, the continuous part of the spectrum and the corresponding eigenstates can also be depicted by the non-Bloch Hamiltonian through the amoeba formulation\cite{37}. The general form of a $d$-dimensional non-Hermitian Hamiltonian with $r$ internal degrees in a unit cell is $H_d = \sum_{\mathbf{n}, \mathbf{l}}  c_{\mathbf{n+l}}^{\dagger} h_{\mathbf{l}} c_{\mathbf{n}}$,
    where $\mathbf{n} = (n_1, n_2 , \cdots, n_d)$ denotes the position of the unit cell with
    $n_i \in [1,L_i]$~($i=1,2, \cdots, d$), $\mathbf{l} = (l_1, l_2 , \cdots, l_d) $ characterizes the hopping range  with $l_i \in [-p_i,q_i]$~($i=1,2, \cdots, d$), and
    $h_{\mathbf{l}}$ is the $r \times r$ hopping matrix. The non-Bloch Hamiltonian corresponding to $H_d$ is
      \begin{equation}
        \mathcal{H}_d (\boldsymbol{\beta})= \mathcal{H}_d (\beta_1, \cdots, \beta_d) = \sum_{\mathbf{l}} h_{\mathbf{l}} \left( \prod_{i=1}^{d} \beta_{i}^{- l_i} \right),
        \label{28}
      \end{equation}
    where $\boldsymbol{\beta} = (\beta_1, \cdots, \beta_d)$ {\zb with} $\beta_i = e^{i \theta_i + \mu_i}$, and $\mu_i$ is the decaying factor in the $i$th direction. The characteristic polynomial of $\mathcal{H}_d (\boldsymbol{\beta})$ is also similarly given by
      \begin{equation}
        f_d (\boldsymbol{\beta},E)  = \det \left[ \mathcal{H}_d (\boldsymbol{\beta}) -  E\mathbb{I}_{r \times r}  \right],
        \label{29}
      \end{equation}
    which can also be expanded as
      \begin{equation}
          f_d (\boldsymbol{\beta}, E) = \sum_{m_j=-r q_j }^{r p_j} \alpha_{j, m_j} \beta_{j}^{m_j}.
         \label{30}
      \end{equation}
    {\wsx
    According to the amoeba formulation, for a fixed $E$, the value of $|\beta_j|$ is determined by
      \begin{equation}
        \nu_j= \oint_{ C_j} \frac{\mathsf{d} \beta_j}{2 \pi i} \frac{\partial_{\beta_j} f_d (\boldsymbol{\beta},E)}{f_d (\boldsymbol{\beta},E)}   = 0 ,
        \label{31}
      \end{equation}
    where $C_j$ is a circle whose radius is $|\beta_j|$, and if $E$ belongs to the OBC spectrum, the range of the value of $|\beta_j|$ reduces to a point\cite{37}.}
     Similar to the 1D case, we find that if $\exists j \in \{1,2,\cdots,d\}$, such that the coefficients of the characteristic polynomial satisfies
      \begin{equation}
        \alpha_{j, - r q_j} = \cdots = \alpha_{j, -1} = 0,
        \label{32}
      \end{equation}
    or
      \begin{equation}
        \alpha_{j, 1} = \cdots = \alpha_{j, r p_j} = 0,
        \label{33}
      \end{equation}
     {\wsx then $|\beta_j| \equiv 0$ or $\infty$, and accordingly} the system is at an IP (see details of the proof in the Supplemental Material\cite{supplemental}). In other words, the criteria for the presence of IPs in a $d$-dimensional non-Hermitian system are given by Eq.\eqref{31} and Eq.\eqref{32}. Apparently, the results in this section is consistent with those in 1D cases. The validity of the criteria given in  Eq.\eqref{31} and Eq.\eqref{32} is exemplified in a 2D non-Hermitian system in the Supplemental Material\cite{supplemental}.
     \par

       Before ending, it is worth remarking that the defectiveness of eigenstates at one IP
      in a higher-dimensional non-Hermitian systems depends on the number of
      $j\in \{1,2,\cdots,d\}$ simultaneously satisfying Eq.\eqref{31} or Eq.\eqref{32}. If
      the number is $m$, the defectiveness will scale as $L^{m}$ in the thermodynamic limit.
      This suggests that IPs in higher-dimensional non-Hermitian systems may have even richer properties.

    \emph{Discussions and conclusions}---We showed that the presence of IPs is connected to the contraction (expansion) of the GBZ to the origin (infinity). In terms of the NBBT, we established the criteria for the presence of IPs in 1D non-Hermitian systems. By further applying the amoeba formulation, we obtained the general criteria for the presence of IPs in higher-dimensional non-Hermitian systems. Our theory solidifies the GBZ-collapse origin of the IPs and provides a very efficient way to determine the condition for the presence of IPs in concrete systems. As applications,
    our theory can be useful for the design of systems supporting IPs and instructive for the study of
    other exotic phases, such as non-Hermitian infernal rings.
    \par

    {\wsx In experiments, platforms utilized to realize non-Hermitian skin effects, such as acoustic systems\cite{Zhang2021winding,Zhou2023geometry} and ultracold atomic systems\cite{41}, which are of high flexibility and controllability, 
    can be used to design models supporting IPs and observe the concomitant ultralocalization of states. Moreover, based on the dynamic skin effect\cite{9,DNHSE2024}, it is also promising to detect the effect induced by IPs through the quantum walk.}

    \emph{Acknowledgments}---This work is supported by the National Natural Science Foundation of China (Grant No. 12174455) and the Natural Science Foundation of Guangdong Province (Grant No. 2021B1515020026).

   \bibliography{defectiveness}
   
   \clearpage

   \onecolumngrid
    \begin{center}
     {\Large  $ \boldsymbol{\mathbf{Supplemental \ Materials}}$}
    \end{center}
    \setcounter{figure}{0}
    \renewcommand\thefigure{S\arabic{figure}}
   \renewcommand\thesection{\Roman{section}}

   \section{Review of the non-Bloch band theory}
   In this section, we give a brief review of the non-Bloch band theory~(NBBT) for one-dimensional (1D) and higher dimensional non-Hermitian systems.
   \par

   Consider a general 1D non-Hermitian system under open boundary conditions (OBCs),
     \begin{equation}
       H = \sum_{n=1}^{L} \sum_{l=-p}^{q} \mathbf{c}_{n+l}^{\dagger} h_l \mathbf{c}_{n},
       \tag{S1}
       \label{S1}
     \end{equation}
   where $L$ is the length of the system, $l$ is the hopping range, $h_l$ is a $r \times r$ matrix,
     \begin{equation}
       \mathbf{c}_{n} = (c_{n,1}, c_{n,2}, \cdots, c_{n,r})^T
       \tag{S2}
       \label{S2}
     \end{equation}
   is the annihilation operator at the $n$th unit cell with $r$ internal degrees, and $\mathbf{c}_{n}^{\dagger}$ is the creation operator corresponding to $\mathbf{c}_{n}$. The non-Bloch Hamiltonian of this system is
     \begin{equation}
       \mathcal{H} (\beta) = \sum_{l=-p}^{q} h_l \beta^{-l},
       \tag{S3}
       \label{S3}
     \end{equation}
   which has $r$ eigenvalues $E_{\mu} (\beta)$~($\mu = 1,2,\cdots,r$) with eigenvectors $\phi_{\mu} (\beta)$~($\mu = 1,2,\cdots,r$). The characteristic polynomial of $H (\beta)$ is
     \begin{equation}
       \begin{split}
         f(\beta,E) = \det \left[ \mathcal{H} (\beta) - E\mathbb{I}_{r \times r} \right]
                    = \sum_{m=-rq}^{rp} \alpha_{m} \beta^m  ,
       \end{split}
       \tag{S4}
       \label{S4}
     \end{equation}
   where $\alpha_m$ are the coefficients whose values depend on the hopping amplitudes and $E$, and $\mathbb{I}_{r \times r}$ is the $r \times r$ identity matrix. $f(\beta,E)$ has $r(p+q)$ roots in total, which can be arranged in the following order,
     \begin{equation}
       |\beta_1 (E)| \leqslant |\beta_2 (E)| \leqslant \cdots \leqslant |\beta_{r(p+q)} (E)|.
       \tag{S5}
       \label{S5}
     \end{equation}
   According to the NBBT\cite{33,34}, the condition
     \begin{equation}
       |\beta_{rq} (E)| = |\beta_{rq+1} (E)|
       \tag{S6}
       \label{S6}
     \end{equation}
   gives the continuous part of the spectrum under OBCs in the thermodynamic limit. All $\beta_{rq} (E)$ and $\beta_{rq+1} (E)$ satisfying Eq.\eqref{S6} compose the GBZ of the system. Generally, the GBZ has $r$ parts, i.e., GBZ$_{\mu}$~($\mu = 1,2,\cdots,r$), which corresponds to the band $E_{\mu} (\beta)$~($\mu = 1,2,\cdots,r$). The GBZ$_{\mu}$ is also called the $\mu$th sub-GBZ. All these sub-GBZs are closed curves in the complex plane, with the origin of the complex plane lying inside them\cite{36,53}. For $\beta \in$GBZ$_{\mu}$, $E_{\mu} (\beta)$ belongs to the continuous part of the spectrum under OBCs, and the wave function of the eigenstate corresponding to $E_{\mu} (\beta)$  has the form
     \begin{equation}
       \Phi_n (E_{\mu} (\beta)) \simeq \beta^n \phi_{\mu} (\beta)
       \tag{S7}
       \label{S7}
     \end{equation}
   in the deep bulk\cite{36}, where $n$ is the position of the unit cell.
   \par

 For higher dimensional non-Hermitian systems under OBCs, the spectrum and the localization property of the skin modes can also be described by the NBBT\cite{37}. The general form of the Hamiltonian for a $d$-dimensional non-Hermitian system with $r$ internal degrees in every unit cell is
   \begin{equation}
     H_d = \sum_{\mathbf{n}, \mathbf{l}}  c_{\mathbf{n+l}}^{\dagger} h_{\mathbf{l}} c_{\mathbf{n}},
     \tag{S8}
     \label{S8}
   \end{equation}
 where $\mathbf{n} = (n_1, n_2 , \cdots, n_d)$ with $n_i \in [1,L_i]$~($i=1,2, \cdots, d$) denotes the position of the unit cell, $\mathbf{l} = (l_1, l_2 , \cdots, l_d) $ with $l_i \in [-p_i,q_i]$~($i=1,2, \cdots, d$) characterizes the hopping range, $h_{\mathbf{l}}$ is the $r \times r$ hopping matrix, and $c_{\mathbf{n}} = \left( c_{\mathbf{n},1}, c_{\mathbf{n},2}, \cdots, c_{\mathbf{n},r} \right)^T$ with $c_{\mathbf{n},i}$
 the annihilation operator corresponding to the $i$th degree and the unit cell at the position $\mathbf{n}$.
 The non-Bloch Hamiltonian corresponding to $H_d$ is
   \begin{equation}
     \mathcal{H}_d (\boldsymbol{\beta})= \mathcal{H}_d (\beta_1, \cdots, \beta_d) = \sum_{\mathbf{l}} h_{\mathbf{l}} \left( \prod_{i=1}^{d} \beta_{i}^{- l_i} \right),
     \tag{S10}
     \label{S10}
   \end{equation}
 where $\boldsymbol{\beta} = (\beta_1, \cdots, \beta_d)$, and $\beta_i = e^{i \theta_i + \mu_i}$ with $\mu_i$ the decaying factor in the $i$th direction.  The continuous spectrum of the system under OBCs can also be obtained by determining the eigenvalues of $\mathcal{H}_d (\boldsymbol{\beta})$ with $\boldsymbol{\beta}$ belonging to the GBZ of the system.
 \par

 The GBZ of higher dimensional non-Hermitian systems can be determined by the amoeba formulation\cite{37}. Consider the characteristic polynomial of $\mathcal{H}_d (\boldsymbol{\beta})$,
   \begin{equation}
     f_d (\boldsymbol{\beta},E)  = \det \left[ \mathcal{H}_d (\boldsymbol{\beta}) -  E\mathbb{I}_{r \times r}  \right].
     \tag{S11}
     \label{S11}
   \end{equation}
 The Ronkin function of the system is defined as
   \begin{equation}
     R \left(E, \mathcal{H}_d, \boldsymbol{\mu}\right) = \int_{T^d} \left(\frac{\mathsf{d} \theta}{2 \pi}\right)^d \ln f_d (\boldsymbol{\beta},E),
     \tag{S12}
     \label{S12}
   \end{equation}
 where $T^d = [0,2\pi]^d$ denotes the $d$-dimensional torus, and $\boldsymbol{\mu} = (\mu_1, \cdots, \mu_d)$. According to the amoeba formulation, if $E$ belongs to the continuous spectrum under OBCs, $R \left(E, \mathcal{H}_d, \boldsymbol{\mu}\right)$ takes the minimum value at unique point
   \begin{equation}
     \boldsymbol{\mu} = \boldsymbol{\mu_0} =  (\mu_{0,1}, \cdots, \mu_{0,d}).
     \tag{S13}
     \label{S13}
   \end{equation}
 That means
   \begin{equation}
     \left. \frac{\partial R \left(E, \mathcal{H}_d, \boldsymbol{\mu}\right)}{\partial \mu_j} \right|_{\boldsymbol{\mu} = \boldsymbol{\mu_0}} = \left. \int_{T^d} \left(\frac{\mathsf{d} \theta}{2 \pi}\right)^d \frac{\partial}{\partial \mu_j} \ln f_d (\boldsymbol{\beta},E) \right|_{\boldsymbol{\mu} = \boldsymbol{\mu_0}} = 0.
     \tag{S14}
     \label{S14}
   \end{equation}
 Since $\beta_j = e^{i \theta_j + \mu_j}$, if one fixes $\mu_j =\mu_{0,j}$ in Eq.\eqref{S14},
   \begin{equation}
     \frac{\partial}{\partial \mu_j} = \beta_j \frac{\partial}{\partial \beta_j}, \qquad
     \mathsf{d} \theta_j = \frac{1}{i \beta_j} \mathsf{d} \beta_j.
     \tag{S15}
     \label{S15}
   \end{equation}
 Thus,
   \begin{equation}
       \left. \int_{ 0}^{2 \pi} \frac{\mathsf{d} \theta_j}{2 \pi} \frac{\partial}{\partial \mu_j} \ln f_d (\boldsymbol{\beta},E) \right|_{\boldsymbol{\mu} = \boldsymbol{\mu_0}} = \left. \oint_{ C_j} \frac{\mathsf{d} \beta_j}{2 \pi i} \frac{\partial_{\beta_j} f_d (\boldsymbol{\beta},E)}{f_d (\boldsymbol{\beta},E)} \right|_{\boldsymbol{\mu} = \boldsymbol{\mu_0}} = \nu_j,
       \tag{S16}
     \label{S16}
   \end{equation}
 where $C_j$ is a circle whose radius is $e^{\mu_{0,j}}$. Apparently, $\nu_j$ is a winding number and the value of $\nu_j$ is a constant. Based on this, Eq.\eqref{S14} can be reduced as
   \begin{equation}
     \left. \frac{\partial R \left(E, \mathcal{H}_d, \boldsymbol{\mu}\right)}{\partial \mu_j} \right|_{\boldsymbol{\mu} = \boldsymbol{\mu_0}} = \int_{T^{d-1}} \left(\frac{\mathsf{d} \theta}{2 \pi}\right)^{d-1} \nu_j = \nu_j =0.
     \tag{S17}
     \label{S17}
   \end{equation}
 In other words, if $E$ belongs to the continuous spectrum under OBCs, for $\forall j=1,2,\cdots,d$, the winding number
   \begin{equation}
     \nu_j = \frac{1}{2 \pi i} \oint_{C_j} \mathsf{d} \beta_j \frac{\partial_{\beta_j} f_d (\boldsymbol{\beta},E)}{f_d (\boldsymbol{\beta},E)} = 0 .
     \tag{S18}
     \label{S18}
   \end{equation}
  According to the amoeba formulation, the asymptotic behavior of the eigenstate corresponding to $E$ in the deep bulk is\cite{37}
   \begin{equation}
     |\Phi_{\mathbf{n}} (E)| \simeq \prod_{i=1}^d e^{\mu_{0,i} n_i}.
     \tag{S19}
     \label{S19}
   \end{equation}

 \section{Fixed point of the $\Lambda$ matrix and the position of the truncation}

   In this section, we review the method in Ref.\cite{32} for calculating the characteristic polynomial of the Hamiltonian under OBCs and show that the fixed point of the $\Lambda$ matrix is consistent with the position of the truncation of the wave function via the Hatano-Nelson (HN) model and the Su-Schrieffer-Heeger (SSH) model in the main text.
   \par

   For the 1D non-Hermitian Hamiltonian given by Eq.\eqref{S1} with $p=q=1$, we can define
     \begin{equation}
        \tilde{H}(j) :=
          \begin{pmatrix}
           \tilde{h}_0  &  h_{-1} &       &      &        \\
           h_1  &  \tilde{h}_0   &  \ddots  &     &      \\
              &   \ddots & \ddots & \ddots &      \\

                &   &    \ddots  &  \tilde{h}_0   &  h_{-1}    \\
                &         &           &  h_1    &   \tilde{h}_0
          \end{pmatrix}_{rj \times rj} ,
          \tag{S20}
          \label{S20}
     \end{equation}
   where $\tilde{h}_0 = h_0 - E \mathbb{I}_{r \times r}$. If the length of the system is $L$, the characteristic polynomial of this system is
     \begin{equation}
        f (E) = \det [\tilde{H}(L)].
        \tag{S21}
        \label{S21}
     \end{equation}
   Next, we can define the initial value and the recursive formula of the $\Lambda$ matrix,
     \begin{equation}
        \Lambda^{(0)} = \tilde{h}_0,
        \quad
        \Lambda^{(1)} = \tilde{h}_0 - h_{-1}  \tilde{h}_{0}^{-1} h_1,
        \quad
        \Lambda^{(j)} = \tilde{h}_0 - h_{-1}  \left(\Lambda^{(j-1)}\right)^{-1} h_1.
        \tag{S22}
        \label{S22}
     \end{equation}
   According to Ref.\cite{32},
     \begin{equation}
       f (E) = \det [\tilde{H}(L)] = \det [\tilde{h}_0] \prod_{j=1}^{L-1} \det [\Lambda^{(j)}].
       \tag{S23}
       \label{S23}
     \end{equation}
   Hence, the spectrum of the system is given by the eigenvalues of $\tilde{h}_0$ and $\Lambda^{(j)}$~($j =1,2,\cdots,L-1$). If the recursive formula has a fixed point, i.e., $\exists j_0$ such that if $j>j_0$, $\Lambda^{(j)} = \Lambda^{(j_0)}$, then the characteristic polynomial has the form
     \begin{equation}
       f (E) = \det [\tilde{H}(L)] = \det [\tilde{h}_0] \left( \det [\Lambda^{(j_0)}] \right)^{L-j_0} \prod_{j=1}^{j_0 -1} \det [\Lambda^{(j)}].
       \tag{S24}
       \label{S24}
     \end{equation}
   In this case, this system has $r(j_0 +1)$ eigenenergies at most, which is independent of $L$. In other words, the spectrum of the system is determined by $\tilde{H}_{j_0+1}$ utterly. This implies that the wave functions of the system have a truncation. Considering the 1D lattice system, because of the discrete translation symmetry, the new system obtained by adding a site at the left of the original system is equal to the one obtained by adding a site at the right of the original system. This means that one cannot identify whether the truncation is at the left side or the right side of the system through this formula. What one can identify is that the wave functions deep in the bulk should have
   the following property,
     \begin{equation}
        \varphi_{n}^T = (0,0,\cdots,0)^T,
        \tag{S25}
        \label{S25}
     \end{equation}
   for $n>j_0-1$ or $n< L - j_0 +1$.
   \par

   For the HN model given in Eq.(1) of the main text,
     \begin{equation}
       \tilde{h}_0^{HN} = -E , \qquad  h_{-1}^{HN} = t_L, \qquad   h_1^{HN} = t_R.
       \tag{S26}
       \label{S26}
     \end{equation}
   If $t_R$ = 0, $-E=\Lambda^{(0)}_{HN} =\Lambda^{(1)}_{HN} = \cdots $, which indicates $j_0^{HN} = 0$. Accordingly, the
   components of the unique wave function of the system, $\Psi_L$, are non-zero only at the first site on the left side. If $t_L$ = 0, $-E=\Lambda^{(0)}_{HN} =\Lambda^{(1)}_{HN} = \cdots $, which also indicates $j_0^{HN} = 0$. For this case, the components of the unique wave function of the system, $\Psi_R$, are non-zero only at the first site on the right side.
   \par

   For the non-Hermitian SSH model given in Eq.(11) of the main text,
     \begin{equation}
       \tilde{h}_0^{SSH} =
          \begin{pmatrix}
             -E  &   t_1 + \gamma   \\
             t_1 - \gamma   &   -E
          \end{pmatrix},
       \qquad
       h_{-1} =
         \begin{pmatrix}
            0   &   0\\
            t_2   &  0
         \end{pmatrix},
         \qquad
       h_{1} =
         \begin{pmatrix}
            0   &   t_2 \\
            0   &  0
         \end{pmatrix}.
         \tag{S27}
         \label{S27}
     \end{equation}
   If $t_1 = \gamma$
     \begin{equation}
       \Lambda^{(1)}_{SSH} = \Lambda^{(2)}_{SSH} =
       \begin{pmatrix}
          -E   &   2 \gamma  \\
          0    &  -E + \frac{t_2^2}{E}
       \end{pmatrix}.
       \tag{S28}
       \label{S28}
     \end{equation}
   If $t_1 = -\gamma$
     \begin{equation}
       \Lambda^{(1)}_{SSH} = \Lambda^{(2)}_{SSH} =
       \begin{pmatrix}
          -E   &   0  \\
          -2 \gamma    &  -E + \frac{t_2^2}{E}
       \end{pmatrix}.
       \tag{S29}
       \label{S29}
     \end{equation}
   Apparently, $j_0^{SSH} = 1$. From the explicit forms of the wave functions given in Eq.(12) and Eq.(13) of
   the main text, one can see that the number and
   localization range of the wave functions of this system agree with the truncation given by $j_0^{SSH} = 1$.

 \section{Proof of the criteria for the presence of infernal points in 1D non-Hermitian systems}

   In this section, we give a proof to the criteria given in Eq.(9) and Eq.(10) of the main text.
   \par

   Let us return to the characteristic polynomial given in Eq.(8). The $r(p+q)$ roots of Eq.(8) are obtained by finding the zeros of the polynomial
     \begin{equation}
        g(\beta, E) = \beta^{rq} f(\beta, E) = \sum_{m=0}^{r(p+q)} \alpha_{m-rq} \beta^{m}.
        \tag{S30}
        \label{S30}
     \end{equation}
   Equivalently, let $\tilde{\beta} = \beta^{-1}$, then the $r(p+q)$ roots can be obtained by finding the zeros of the polynomial
     \begin{equation}
       \tilde{g} (\tilde{\beta}, E) = \tilde{\beta}^{rp} f(\tilde{\beta}, E) = \sum_{m=0}^{r(p+q)} \alpha_{-m+rp} \tilde{\beta}^{m}
       \tag{S31}
       \label{S31}
     \end{equation}
   For the case,
     \begin{equation}
        \alpha_m \equiv 0 \qquad (-rq \leqslant m < -m_1 \quad  \mathsf{and} \quad m_2 < m \leqslant rp),
        \tag{S32}
       \label{S32}
     \end{equation}
   where $-rq < -m_1 \leqslant 0 \leqslant m_2 < rp$,
     \begin{gather}
       g(\beta, E) = \beta^{rq-m_1}\left( \sum_{m=0}^{m_2+m_1} \alpha_{m-m_1} \beta^m \right),
       \tag{S33}
       \label{S33}
        \\
       \tilde{g} (\tilde{\beta}, E) = \tilde{\beta}^{rp-m_2} \left( \sum_{m=0}^{m_2+m_1} \alpha_{m_2 - m} \tilde{\beta}^m \right).
       \tag{S34}
       \label{S34}
     \end{gather}
   Apparently, $rq-m_1$ roots of Eq.\eqref{S33} and $rp-m_2$ roots of Eq.\eqref{S34} are $0$, i.e.,
     \begin{gather}
       |\beta_1| \equiv |\beta_2| \equiv \cdots \equiv |\beta_{rq-m_1}| \equiv 0 \leqslant |\beta_{rq+m_1+1}|  \leqslant \cdots \leqslant |\beta_{r(p+q)}|,
       \tag{S35}
       \label{S35}
        \\
       |\tilde{\beta}_1| \equiv |\tilde{\beta}_2| \equiv \cdots \equiv |\tilde{\beta}_{rp-m_2}| \equiv 0 \leqslant |\tilde{\beta}_{rp-m_2+1}|  \leqslant \cdots \leqslant |\tilde{\beta}_{r(p+q)}|.
       \tag{S36}
       \label{S36}
     \end{gather}
   Since $\tilde{\beta} = \beta^{-1}$, for the $r(p+q)$ roots of Eq.(8) under the constraints given by Eq.\eqref{S32},
     \begin{gather}
       |\beta_1| \equiv |\beta_2| \equiv \cdots \equiv |\beta_{rq+m_1}| \equiv 0 ,
       \tag{S37}
       \label{S37}
       \\
       |\beta_{r(p+q)}| \equiv |\beta_{r(p+q)-1}| \equiv \cdots \equiv |\beta_{rq+m_2+1}| \equiv \infty.
       \tag{S38}
       \label{S38}
     \end{gather}
   Combining Eq.\eqref{S37}, Eq.\eqref{S38} with Eq.\eqref{S6}, one finds that if $m_1 = 0$, the condition given in Eq.\eqref{S6} can be reduced as $|\beta_{rq}|=|\beta_{rq+1}| = 0$, which corresponds to the contraction of the GBZ to the origin of the complex plane.
   On the other hand, if $m_2 = 0$, the condition given in Eq.\eqref{S6} can be reduced as $|\beta_{rq}|=|\beta_{rq+1}| = \infty$, which
    corresponds to the expansion of the GBZ to the infinity of the complex plane. As elucidated in the main text, the collapse of the GBZ leads to presence of infernal points. Hence, the criteria for the presence of infernal points in 1D non-Hermitian systems are given by Eq.(18) and Eq.(19) in the main text.
   \par

 \section{Proof of the criteria for the presence of infernal points in higher-dimensional non-Hermitian systems}

   In this section we give a proof to the criteria of the presence of infernal points in higher-dimensional non-Hermitian systems based on the amoeba formulation.
   \par

   For a $d$-dimensional non-Hermitian systems with $d\geq2$, the asymptotic behavior of the wave functions in the deep bulk has the form given in Eq.\eqref{S19}. This implies that if $\beta_j \equiv 0$ or $\beta_j \equiv \infty$ for the GBZ, all wave functions will have a truncation in the $j$th direction. Similar to the 1D case, the collapse of $\beta_j$ for GBZ leads to the presence of infernal points. Thus, we only need to find conditions for $\beta_j \equiv 0$ or $\beta_j \equiv \infty$~($j \in \{ 1,2, \cdots, d \}$).
   \par

   We start with the characteristic polynomial, $f_d (\boldsymbol{\beta},E)$, given in Eq.(19), which is the characteristic polynomial for a general $d$-dimensional non-Bloch Hamiltonian. Expanding $f_d (\boldsymbol{\beta},E)$ with respect to the variable $\beta_j$, we get
     \begin{equation}
       f_d (\boldsymbol{\beta}, E) = f_d (\beta_j) = \sum_{m_j=-r q_j }^{r p_j} \alpha_{j, m_j} \beta_{j}^{m_j}
       \tag{S39}
       \label{S39}
     \end{equation}
   which is Eq.(20) in the main text. According to the definition of $\nu_j$ in Eq.\eqref{S16},
     \begin{equation}
       \nu_j = \# N - \# P,
       \tag{S40}
       \label{S40}
     \end{equation}
   where $\# N$ and $\# P$ represent the number of zeros and poles lying inside the curve $C_j$, respectively. It is noteworthy that an $m$ order zero (pole) is counted as $m$ zeros (poles). If $\alpha_{j, - r q_j} = \cdots = \alpha_{j, -1} = 0$, $f_d (\beta_j)$ has no pole. In this case, zeros of $f_d (\beta_j)$ are denoted as $\beta_{j,1}, \beta_{j,2}, \beta_{j,3},\cdots, \beta_{j,r p_j}$, and
     \begin{equation}
        |\beta_{j,1}| \leqslant |\beta_{j,2}| \leqslant |\beta_{j,3}| \leqslant \cdots \leqslant |\beta_{j,r p_j}|  .
        \tag{S41}
        \label{S41}
     \end{equation}
   Thus, to have $\nu_j = 0$, one must take $0 \leqslant e^{\mu_{j}} <|\beta_{j,1}| $. According to the amoeba formulation\cite{37}, if $E$ belongs to the continuous spectrum of the system under OBCs, the value range for $e^{\mu_{j}}$ that makes $\nu_j = 0$ will degenerate to a point. That means for all eigenstates of the system, the value of $\beta_j$ corresponding to them is $0$, i.e., $|\beta_j| \equiv 0$.

   On the other hand, if $\alpha_{j, 1} = \cdots = \alpha_{j, r p_j} = 0$, $f_d (\beta_j)$ can be written as
     \begin{equation}
       f_d (\beta_j) = \beta_j^{-r q_j} \sum_{m_j=-r q_j }^{0} \alpha_{j, m_j} \beta_{j}^{m_j + r q_j}.
       \tag{S42}
       \label{S42}
     \end{equation}
   Hence, $f_d (\beta_j)$ has a $r q_j$ order pole at $\beta_j = 0$ and $r q_j$ zero points denoted as
     \begin{equation}
       |\beta_{j,1}| \leqslant |\beta_{j,2}| \leqslant |\beta_{j,3}| \leqslant \cdots \leqslant |\beta_{j,r q_j}|,
       \tag{S43}
       \label{S43}
     \end{equation}
   in this case. Thus, to have $\nu_j = 0$, the value range of $e^{\mu_{j}}$ needs to satisfy $|\beta_{j,r q_j}| < e^{\mu_{j}} \leqslant \infty$. If $E$ belongs to the continuous spectrum of the system under OBCs, this range will expand to the infinity of the complex plane, i.e., $|\beta_j| \equiv \infty$ in this case. These results lead to the criteria (Eq.(21) or Eq.(22) in the main
   text) for the presence of infernal points in higher-dimensional non-Hermitian systems.
   \par

   If there are $m$ directions ($1\leq m\leq d$) simultaneously satisfying $\alpha_{j, - r q_j} = \cdots = \alpha_{j, -1} = 0$ or
   $\alpha_{j, 1} = \cdots = \alpha_{j, r p_j} = 0$, then $m$ components of ($\beta_{1},\beta_{2},...,\beta_{d})$ are degenerate,
   and the wave functions in real space are truncated in $m$ directions. This means that the eigenenergies and eigenstates of the system can be fully determined by a truncated Hamiltonian whose matrix size scales as $L^{d-m} \times L^{d-m}$
   in the thermodynamic limit. In this case, the defectiveness of eigenstates at an IP will scale as  $L^{m}$.

 \section{Examples}
   In this section, we provide three more examples to show the validity and generality of the criteria. The first example is the generalized non-Hermitian SSH model whose GBZ is no longer a circle in general. The second example is a 1D non-Hermitian system in the symplectic class. The third example is a 2D non-Hermitian system.

   \subsection{Generalized non-Hermitian SSH model}
     The Hamiltonian for the generalized non-Hermitian SSH model is
       \begin{equation}
         \begin{split}
           H_{GSSH} = \sum_{n=1}^{L}& (t_1 + \gamma) c_{n,A}^{\dagger} c_{n,B} + (t_1 - \gamma) c_{n,B}^{\dagger} c_{n,A}  \\
                     & + (t_2 + \delta)  c_{n+1,A}^{\dagger} c_{n,B} + (t_2 - \delta) c_{n-1,B}^{\dagger} c_{n,A}  \\
                     & +  (t_3 + \eta)  c_{n-1,A}^{\dagger} c_{n,B} + (t_3 - \eta) c_{n+1,B}^{\dagger} c_{n,A}.
          \end{split}
          \tag{S44}
          \label{S44}
       \end{equation}
     The non-Bloch Hamiltonian corresponding to $H_{GSSH}$ is
       \begin{equation}
         \mathcal{H}_{GSSH} (\beta) = h_{0}^{GSSH} + h_{1}^{GSSH} \beta^{-1} + h_{-1}^{GSSH} \beta,
         \tag{S45}
         \label{S45}
       \end{equation}
     where
       \begin{equation}
         h_{0}^{GSSH} =
           \begin{pmatrix}
              0    &    t_1 + \gamma  \\
             t_1 - \gamma &   0
           \end{pmatrix},
         \quad
         h_{1}^{GSSH} =
           \begin{pmatrix}
             0    &    t_2 + \delta  \\
             t_3 - \eta &   0
           \end{pmatrix},
         \quad
         h_{-1}^{GSSH} =
           \begin{pmatrix}
             0    &    t_3 + \eta  \\
             t_2 - \delta &   0
           \end{pmatrix}.
         \tag{S46}
         \label{S46}
       \end{equation}
       \begin{figure}
         \centering
         \includegraphics[scale=0.5]{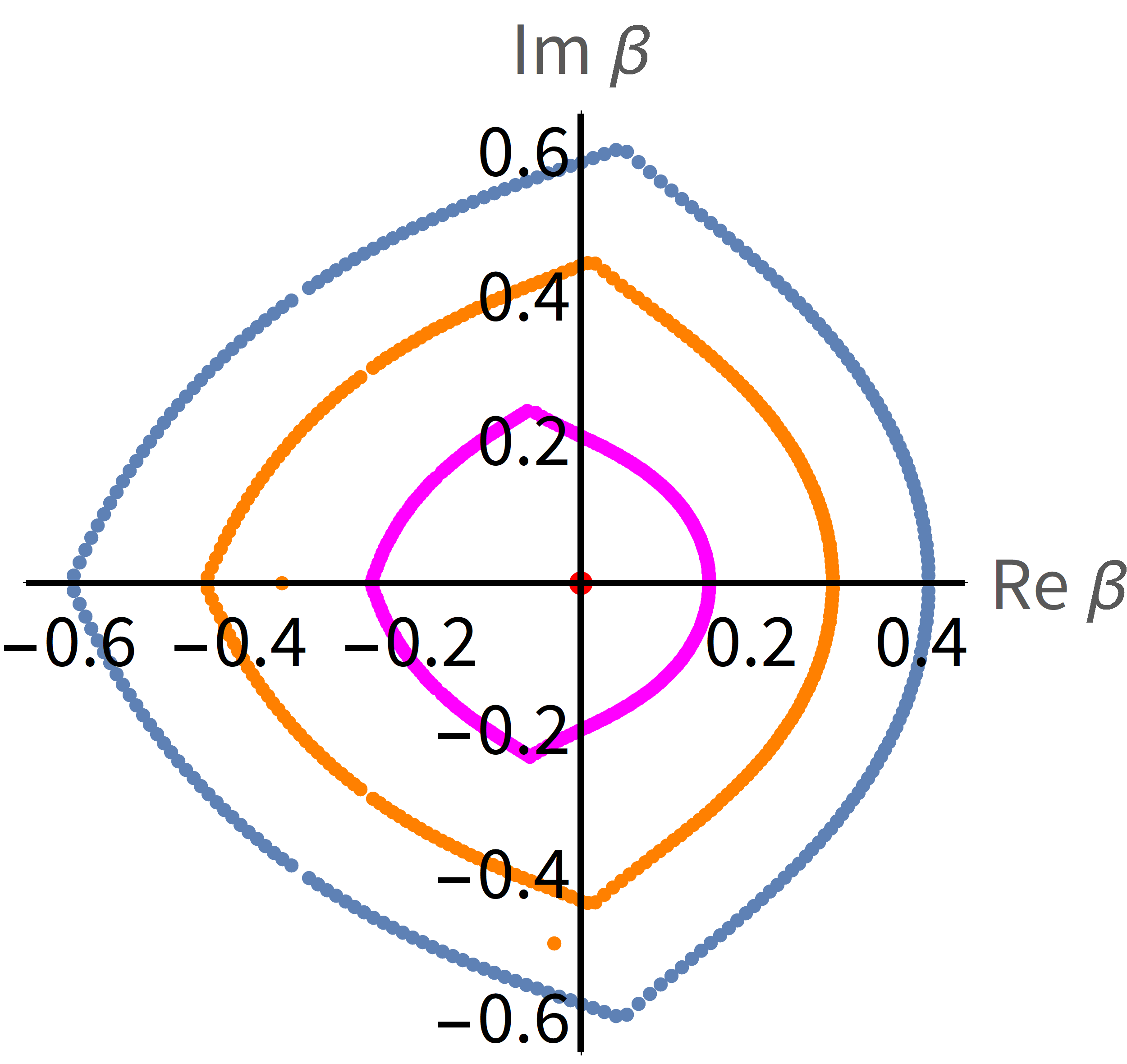}
         \caption{The blue, orange, magenta curve and the red point are GBZ of the generalized non-Hermitian SSH model for $\delta=-1$, $\delta=-2$, $\delta=-\frac{14}{5}$ and $\delta=-3$ respectively. Other parameters are $t_1=1$, $\gamma = \frac{-3}{4}$, $t_2=3$, $t_3=2$, $\eta =2$. The length of the system is $L=120$}
         \label{figs1}
       \end{figure}

     \par

     The characteristic polynomial of $\mathcal{H}_{GSSH} (\beta)$ is
       \begin{equation}
          \begin{split}
             f_{GSSH} (\beta, E) =& \det [\mathcal{H}_{GSSH} (\beta) - E \mathbb{I}_{2 \times 2}]
             \\
              =& -(t_2 + \delta)(t_3 - \eta) \beta^{-2} + \left[ \gamma (t_2 - t_3 + \delta + \eta ) -t_1 (t_2 + t_3 + \delta - \eta) \right] \beta^{-1}
             \\
              &  + \left[ \gamma (-t_2 + t_3 + \delta + \eta ) -t_1 (t_2 + t_3 - \delta + \eta) \right] \beta  -(t_2 - \delta)(t_3 + \eta) \beta^{2}
              \\
              & + E^2 + \gamma^2 + \delta^2 + \eta^2 - t_1^2 - t_2^2 - t_3^2.
          \end{split}
          \tag{S47}
          \label{S47}
       \end{equation}
     According to our theory, it is easy to find two infernal points. One of them is at $t_2 = - \delta$ and $t_3 = \eta$, which corresponds to $|\beta| \equiv 0$, and the other is at $t_2 =  \delta$ and $t_3 = -\eta$, which corresponds to $|\beta| \equiv \infty$.  In Fig.\ref{figs1}, one can see that the GBZ of the system is in general not a circle. If one fixes $t_3 = \eta$ and tunes $\delta$ towards $-t_2$, the GBZ of this system will shrink to the origin of the complex plane eventually.
     \par
     Solving the system directly, we find that if $t_2 = - \delta$ and $t_3 = \eta$, the system  has only two eigenstates whose
     explicit forms read
       \begin{gather}
         \Phi_{L,1}^{GSSH} = \left( \frac{\sqrt{t_1^2 - \gamma^2}}{t_1 - \gamma},1,0,\cdots,0 \right)^T,
         \tag{S48}
         \label{S48}
         \\
         \Phi_{L,2}^{GSSH} = \left( -\frac{\sqrt{t_1^2 - \gamma^2}}{t_1 - \gamma},1,0,\cdots,0 \right)^T,
         \tag{S49}
         \label{S49}
       \end{gather}
     and the eigenenergies corresponding to $\Phi_{L,1}^{GSSH}$ and $\Phi_{L,2}^{GSSH}$ are $\sqrt{t_1^2 - \gamma^2}$ and $-\sqrt{t_1^2 - \gamma^2}$, respectively. On the other hand, if $t_2 =  \delta$ and $t_3 = -\eta$, the system also has only two eigenstates, whose explicit forms are
       \begin{gather}
         \Phi_{R,1}^{GSSH} = \left( 0,0,\cdots,0,\frac{\sqrt{t_1^2 - \gamma^2}}{t_1 - \gamma},1 \right)^T,
         \tag{S50}
         \label{S50}
         \\
         \Phi_{R,2}^{GSSH} = \left( 0,0,\cdots,0,-\frac{\sqrt{t_1^2 - \gamma^2}}{t_1 - \gamma},1 \right)^T,
         \tag{S51}
         \label{S51}
       \end{gather}
     and the eigenenergies corresponding to $\Phi_{L,1}^{GSSH}$ and $\Phi_{L,2}^{GSSH}$ are $\sqrt{t_1^2 - \gamma^2}$ and $-\sqrt{t_1^2 - \gamma^2}$, respectively. For the case $t_2 = - \delta$ and $t_3 = \eta$, $h_{1}^{GSSH} = \mathbf{0}_{2 \times 2}$ and for the case $t_2 =  \delta$ and $t_3 = -\eta$, $h_{-1}^{GSSH} = \mathbf{0}_{2 \times 2}$. Thus, according to the theory given in Sec. $\mathbf{II}$, the value of the $\Lambda$ matrix is
       \begin{equation}
         \begin{pmatrix}
            -E   &   t_1 + \gamma  \\
            t_1 - \gamma  &  -E
         \end{pmatrix}
         = \Lambda^{(0)}_{GSSH} = \Lambda^{(1)}_{GSSH} = \cdots = \Lambda^{(L-1)}_{GSSH},
         \tag{S52}
         \label{S52}
       \end{equation}
     which is the same for both infernal points of the system. Hence, $j_0^{GSSH} = 0$. According to
     the explicit forms of $\Phi_{L,1}^{GSSH}$ and $\Phi_{L,2}^{GSSH}$ given in Eq.(\ref{S48}) and Eq.(\ref{S49}) (or
     $\Phi_{R,1}^{GSSH}$ and $\Phi_{R,2}^{GSSH}$ in Eq.(\ref{S50}) and Eq.(\ref{S51})), again one can see that
     the number and localization range of the wave functions of this system agree with the truncation.
     \par

     {\wsx  It is worth to point out that the existence of infernal points does  not mean the disappearance of hoppings in some directions. In fact, it is not hard to verify that $t_1 = \gamma$ and $t_3 = \eta$ is also an infernal point for the generalized non-Hermitian SSH model, and at this infernal point, it is apparent that the system has hoppings in both directions.}

   \subsection{1D non-Hermitian in the symplectic class}
     In this section, we discuss the impact of symmetry on the criteria for the presence of infernal points.
     We consider the four-band system given in Ref.\cite{31}, whose non-Bloch Hamiltonian is
       \begin{equation}
          \mathcal{H}_{sc} (\beta) = \sum_{i=-1}^{1} h_{i}^{sc} \beta^{-i},
          \tag{S53}
          \label{S53}
       \end{equation}
     where
       \begin{gather}
         h_{0}^{sc} =
           \begin{pmatrix}
              0    &    0    &    M+ \delta    &    0    \\
              0    &    0    &    0    &    M+ \delta    \\
              M- \delta    &    0    &    0    &    0    \\
              0    &   M- \delta    &    0    &    0
           \end{pmatrix},
         \tag{S54}
         \label{S54}
         \\
         h_{1}^{sc} =
           \begin{pmatrix}
             0    &    0    &    \frac{1 + \lambda}{2}    &    0    \\
             0    &    0    &    0    &    \frac{1 - \lambda}{2}    \\
             \frac{1 - \lambda}{2}    &    0    &    0    &    0    \\
             0    &   \frac{1 + \lambda}{2}    &    0    &    0
           \end{pmatrix},
         \tag{S55}
         \label{S55}
         \\
         h_{-1}^{sc} =
           \begin{pmatrix}
             0    &    0    &    \frac{1 - \lambda}{2}    &    0    \\
             0    &    0    &    0    &    \frac{1 + \lambda}{2}    \\
             \frac{1 + \lambda}{2}    &    0    &    0    &    0    \\
             0    &   \frac{1 - \lambda}{2}    &    0    &    0
           \end{pmatrix}.
           \tag{S56}
         \label{S56}
       \end{gather}
     The non-Bloch Hamiltonian, $\mathcal{H}_{sc} (\beta)$, has TRS$^{\dagger}$ ,
       \begin{equation}
          U \mathcal{H}_{sc} (\beta) U^{-1} = \mathcal{H}_{sc}^{T} (\beta^{-1}),
          \tag{S57}
          \label{S57}
       \end{equation}
     where
       \begin{gather}
          U = U^{-1} = \sigma_x \otimes \sigma_x,
          \tag{S58}
          \label{S58}
          \\
          \sigma_x =
            \begin{pmatrix}
               0  &  1  \\
               1  &  0
            \end{pmatrix}.
            \tag{S59}
            \label{S59}
       \end{gather}
     The Hamiltonian hence belongs to the symplectic class.  The characteristic polynomial of $\mathcal{H}_{sc} (\beta)$ is
       \begin{equation}
          \begin{split}
             f_{sc} (\beta,E) = & \det [\mathcal{H}_{sc} (\beta) - E \mathbb{I}_{4 \times 4}] \\
              = & \sum_{i=-4}^{4} \alpha_{i}^{sc} \beta^i.
          \end{split}
          \tag{S60}
          \label{S60}
       \end{equation}
     Without any constraint,  $f_{sc} (\beta,E)$ has $8$ independent roots which can be arranged in the order
       \begin{equation}
          |\beta_1^{sc}| \leqslant |\beta_2^{sc}|  \leqslant |\beta_3^{sc}| \leqslant |\beta_4^{sc}| \leqslant |\beta_5^{sc}| \leqslant |\beta_6^{sc}| \leqslant |\beta_7^{sc}| \leqslant |\beta_8^{sc}|.
          \tag{S61}
          \label{S61}
       \end{equation}
     Because of the existence of TRS$^{\dagger}$, the $8$ roots are no longer independent and satisfy the following relation,
       \begin{equation}
         \beta_1^{sc} = \left(\beta_8^{sc}\right)^{-1}, \quad  \beta_2^{sc} = \left(\beta_7^{sc}\right)^{-1},  \quad  \beta_3^{sc} = \left(\beta_6^{sc}\right)^{-1}, \quad  \beta_4^{sc} = \left(\beta_5^{sc}\right)^{-1}
         \tag{S62}
          \label{S62}.
       \end{equation}
     According to Ref.\cite{35}, the condition for determining GBZ becomes
       \begin{equation}
         |\beta_3^{sc}| = |\beta_4^{sc}|, \qquad  |\beta_5^{sc}| = |\beta_6^{sc}|,
         \tag{S63}
         \label{S63}
       \end{equation}
     and two sub-GBZs are composed by ($\beta_3^{sc}$, $\beta_4^{sc}$) and ($\beta_5^{sc}$, $\beta_6^{sc}$) respectively. We use GBZ$_1$ to denote the sub-GBZ composed by ($\beta_3^{sc}$, $\beta_4^{sc}$) and use GBZ$_2$ to denote the sub-GBZ composed by ($\beta_5^{sc}$, $\beta_6^{sc}$). Due to the constraint given in Eq.\eqref{S62}, if GBZ$_1$ shrinks to the origin of the complex plane, then GBZ$_2$ must expand to the infinity of the complex plane. Thus, the criteria for the collapse of GBZs become
       \begin{equation}
         |\beta_1^{sc}| = |\beta_2^{sc}|  = |\beta_3^{sc}| \equiv 0,
         \qquad
         |\beta_6^{sc}| = |\beta_7^{sc}|  = |\beta_8^{sc}| \equiv \infty.
         \tag{S64}
         \label{S64}
       \end{equation}
     This means that the coefficients in Eq.\eqref{S60} must satisfy
       \begin{equation}
         \alpha_{-4}^{sc} = \alpha_{-3}^{sc} = \alpha_{-2}^{sc} = \alpha_{2}^{sc} = \alpha_{3}^{sc} = \alpha_{4}^{sc} \equiv 0
         \tag{S65}
         \label{S65}
       \end{equation}
     according to Sec. $\mathbf{IV}$. In other words,  TRS$^{\dag}$ will make a connection between
      $\alpha_{m}$ and $\alpha_{-m}$, which accordingly modifies the criterion for
      the presence of infernal points somewhat.
     \par

     For the concerned four-band model, it is not hard to verify that
       \begin{equation}
         f_{sc} (\beta,E) = 2 \delta (E^2 -1) (\beta + \beta^{-1}) + 4 \delta^2 + (E^2 -1)^2
         \tag{S66}
         \label{S66}
       \end{equation}
     when $\lambda = 1$ and $M= -\delta $. Apparently, the values of the coefficients $\alpha_{i}$ satisfy Eq.\eqref{S65}.
     Solving the system directly, we find that the system has only $6$ eigenstates\cite{31}, whose explicit forms read
       \begin{gather}
         \Phi_{L,0}^{sc} = (0,0,0,1,0,\cdots,0)^T,
         \tag{S67}
         \label{S67}
         \\
         \Phi_{R,0}^{sc} = (0,\cdots,0,1,0,0,0)^T,
         \tag{S68}
         \label{S68}
         \\
         \Phi_{L,1}^{sc} = (0,1,0,-2 \delta,0,0,0,1,0,\cdots,0)^T,
         \tag{S69}
         \label{S69}
         \\
         \Phi_{L,-1}^{sc} = (0,-1,0,-2 \delta,0,0,0,1,0,\cdots,0)^T,
         \tag{S70}
         \label{S70}
         \\
         \Phi_{R,1}^{sc} = (0,\cdots,0,0,1,0,1,0,-2\delta,0)^T,
         \tag{S71}
         \label{S71}
         \\
         \Phi_{R,-1}^{sc} = (0,\cdots,0,0,1,0,-1,0,-2\delta,0)^T.
         \tag{S72}
         \label{S72}
       \end{gather}
     The eigenenergy is $0$ for $\Phi_{L,0}^{sc}$ and $\Phi_{R,0}^{sc}$, $1$ for $\Phi_{L,1}^{sc}$ and $\Phi_{R,1}^{sc}$, and
     $-1$ for $\Phi_{L,-1}^{sc}$ and $\Phi_{R,-1}^{sc}$, demonstrating the presence of infernal points
     when the criterion is satisfied.
     \par

   \subsection{An example for 2D non-Hermitian systems}

     To show the validity of the criteria for higher-dimensional non-Hermitian systems, here we consider a single-band 2D system with
     size $L\times L$, whose non-Bloch Hamiltonian is
       \begin{equation}
          \begin{split}
           \mathcal{H}_2  ( \beta_x, \beta_y) =& (t_x + \gamma_x) \beta_x + (t_x - \gamma_x) \beta_x^{-1} + (t_y + \gamma_y) \beta_y + (t_y - \gamma_y) \beta_y^{-1}
           \\
           & + t_1 \beta_x  \beta_y + t_2 \beta_x  \beta_y^{-1} + t_3 \beta_x^{-1}  \beta_y + t_4 \beta_x^{-1}  \beta_y^{-1}
          \end{split}
         \tag{S73}
         \label{S73}
       \end{equation}
     In the following, we will focus on the behavior of $\beta_x$ since the behavior of $\beta_y$ is similar to that of $\beta_x$.
     \par

\begin{figure}
        \centering
        \subfigure[]{\includegraphics[scale=0.23]{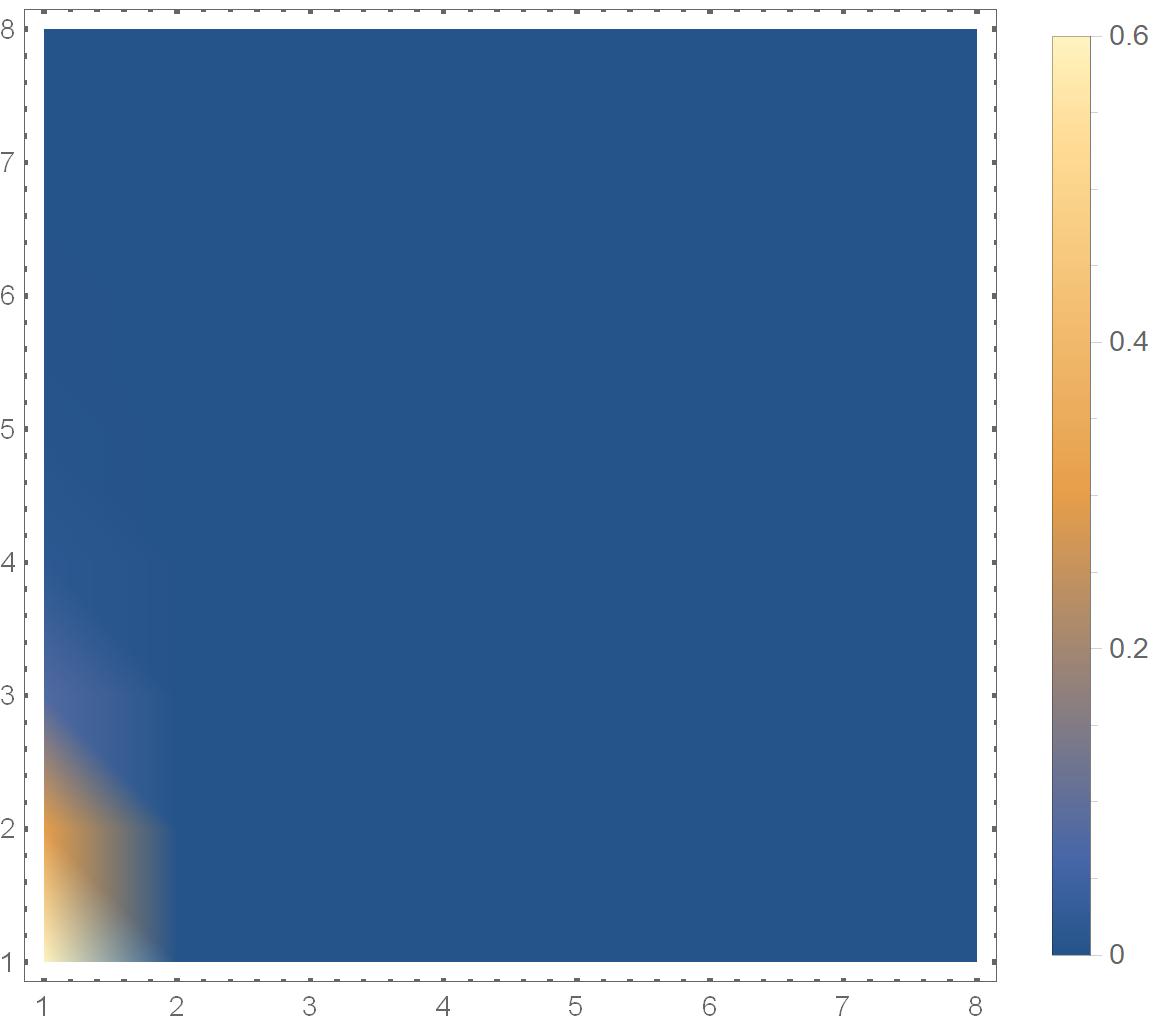} \label{figs2a}}
        \quad
        \subfigure[]{\includegraphics[scale=0.23]{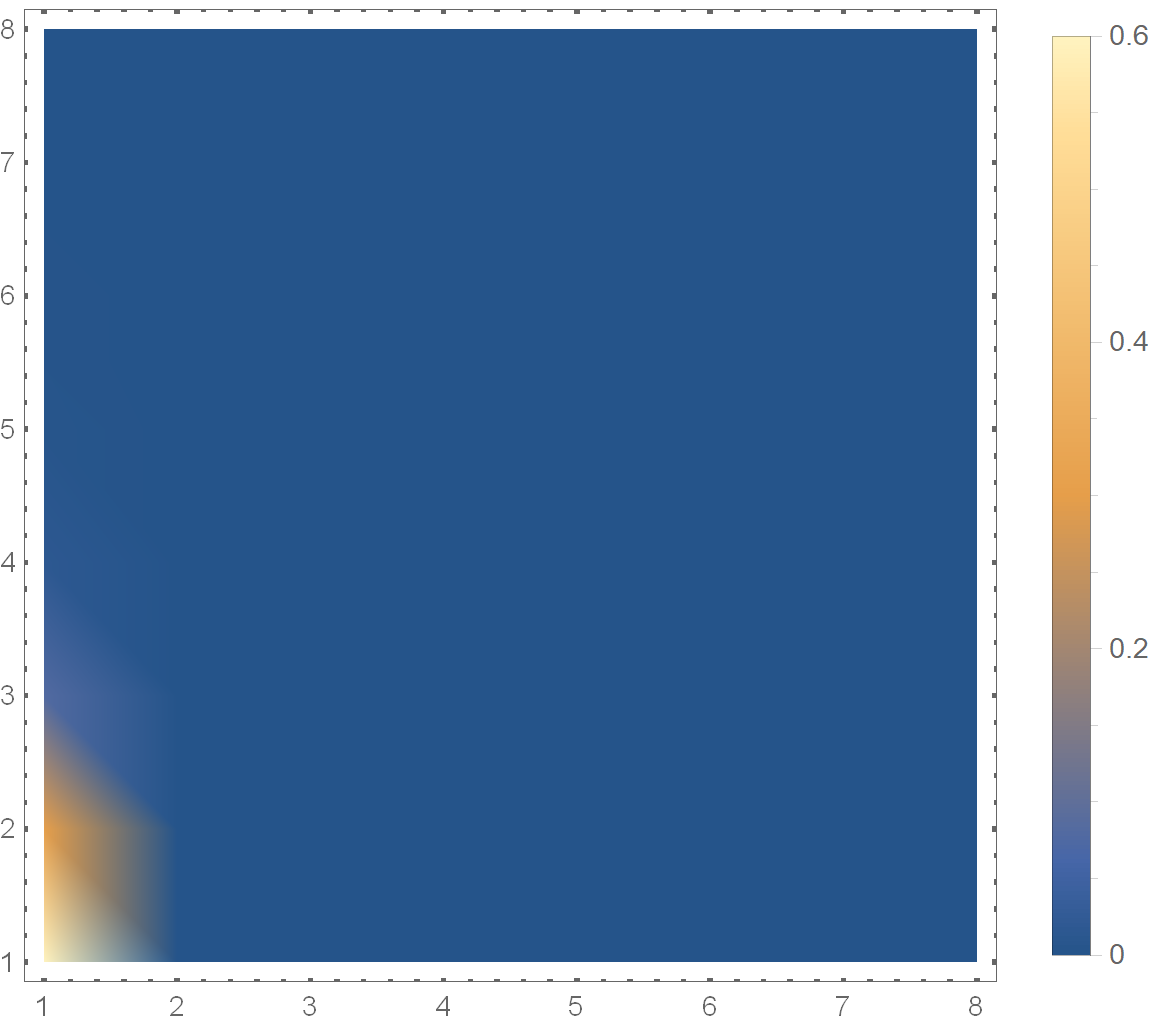} \label{figs2b}}
        \quad
        \subfigure[]{\includegraphics[scale=0.23]{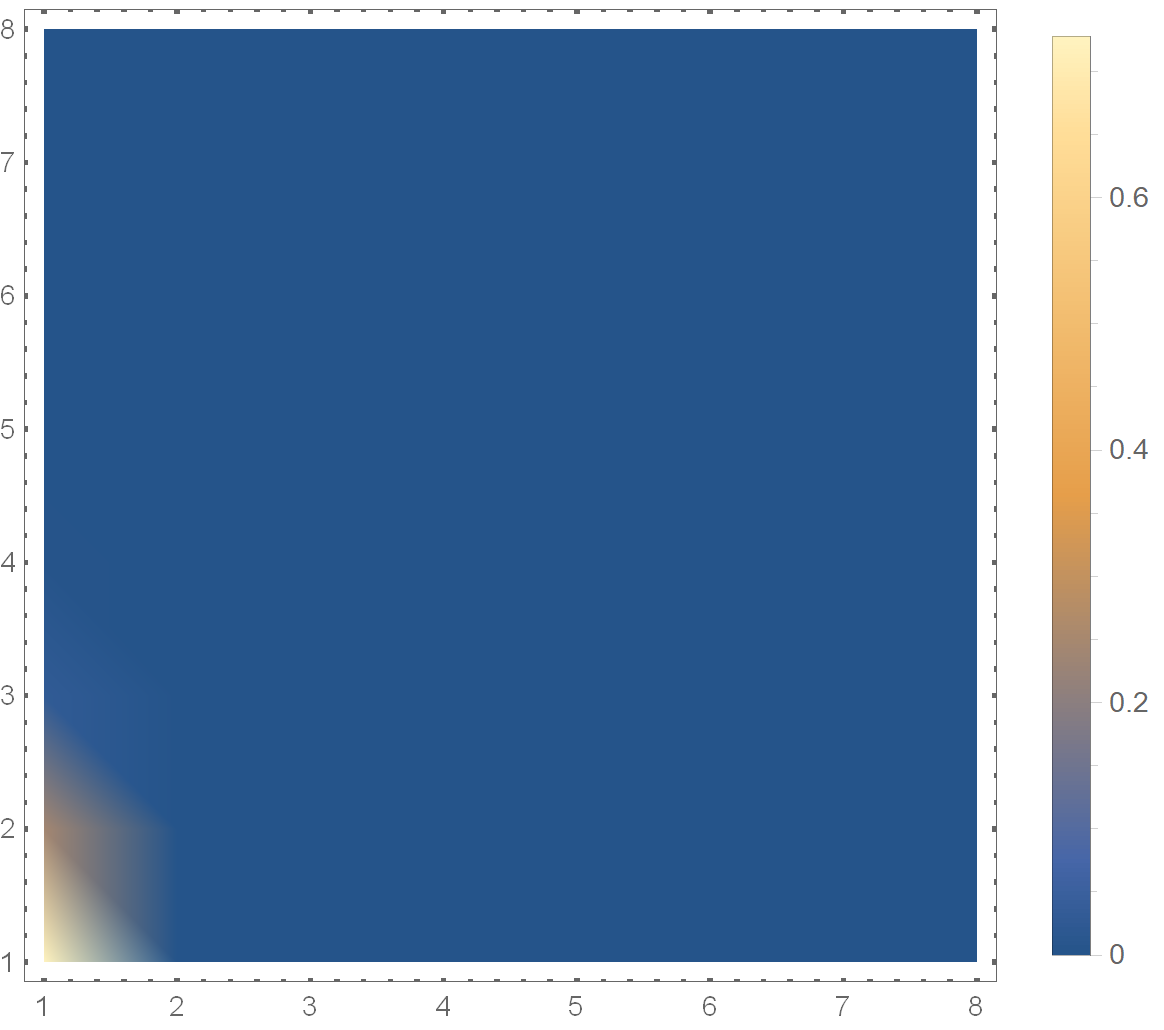} \label{fig2c}}
        \quad
        \subfigure[]{\includegraphics[scale=0.23]{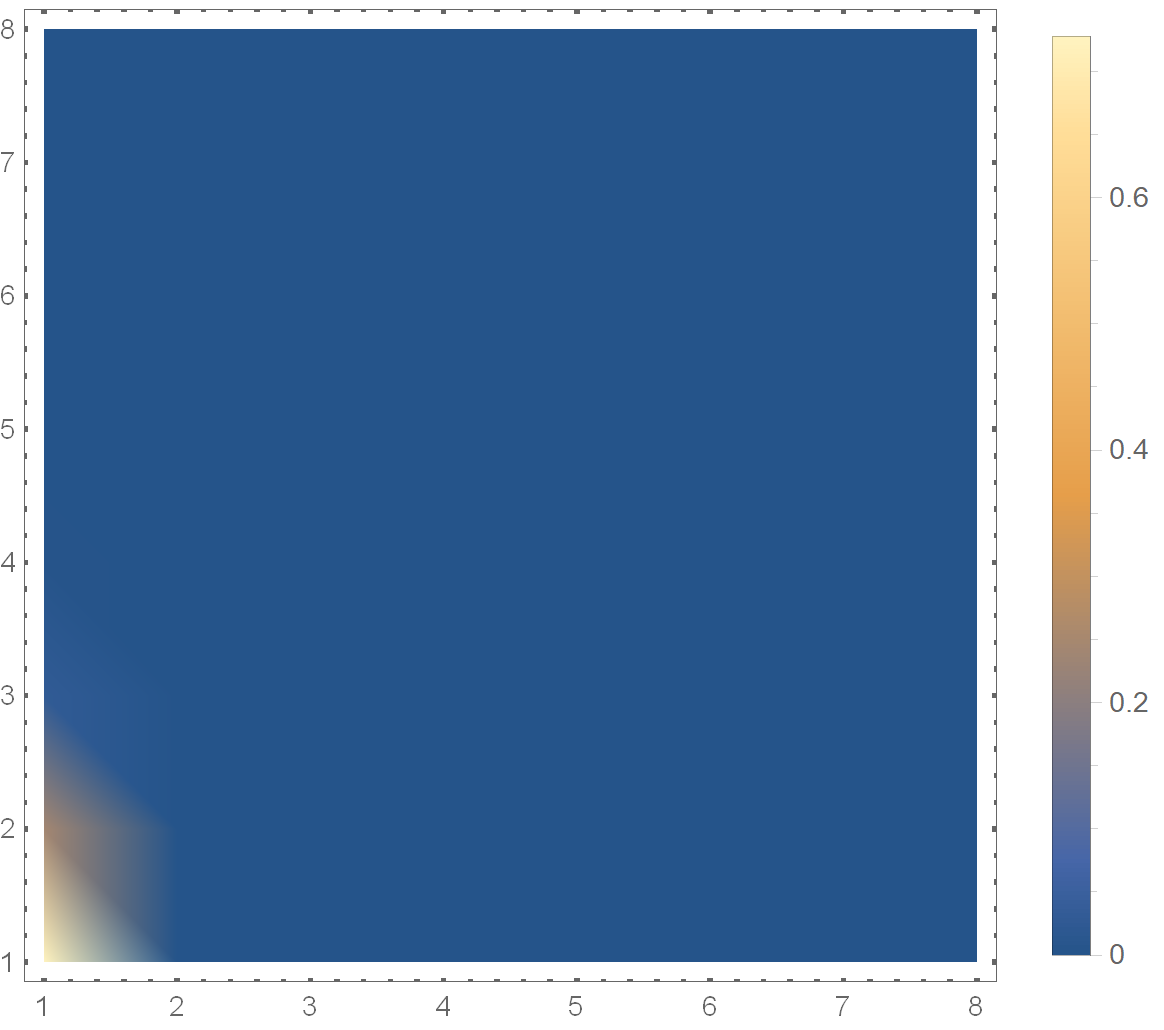} \label{fig2d}}
        \quad
        \subfigure[]{\includegraphics[scale=0.23]{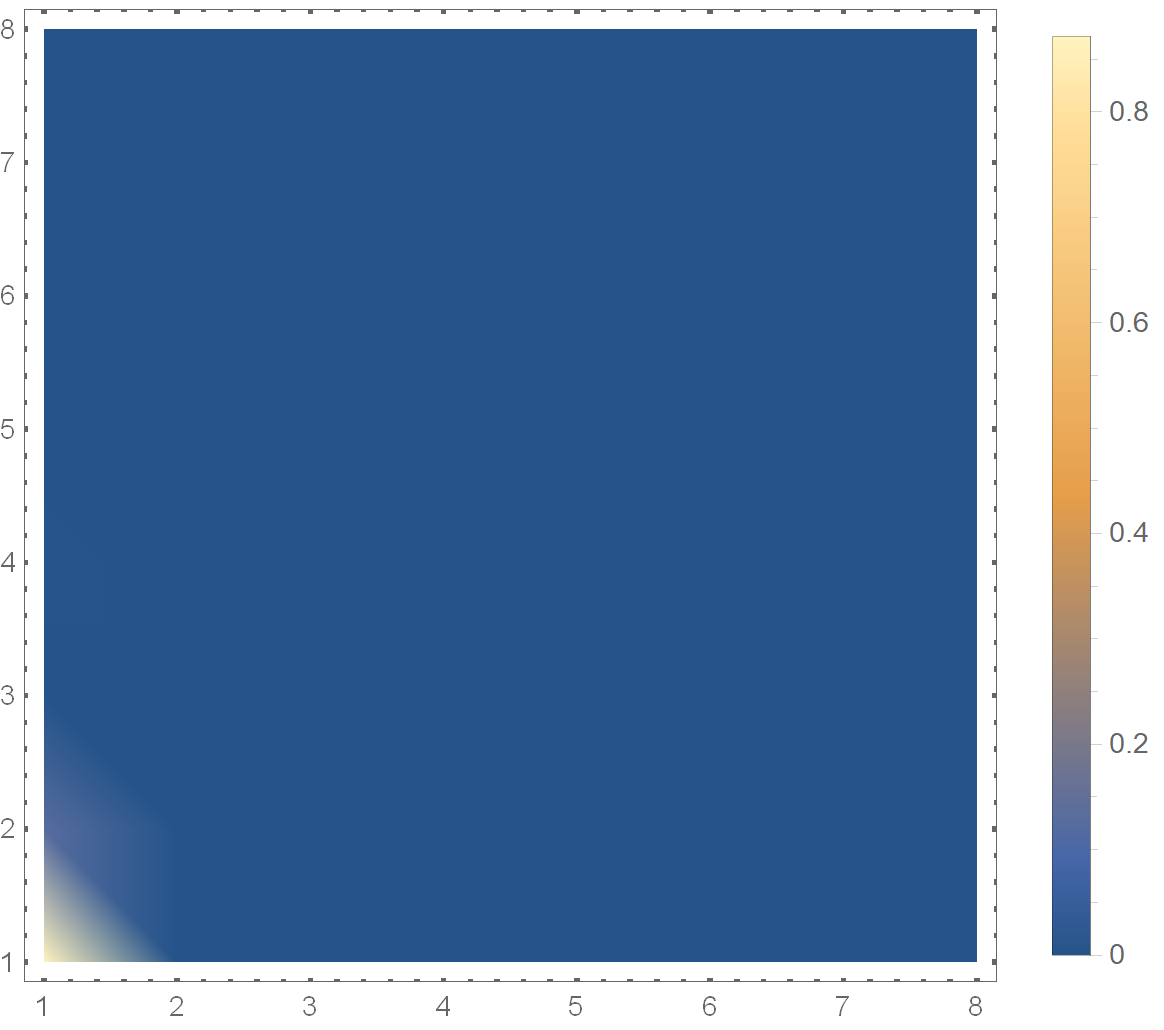} \label{fig2e}}
        \quad
        \subfigure[]{\includegraphics[scale=0.23]{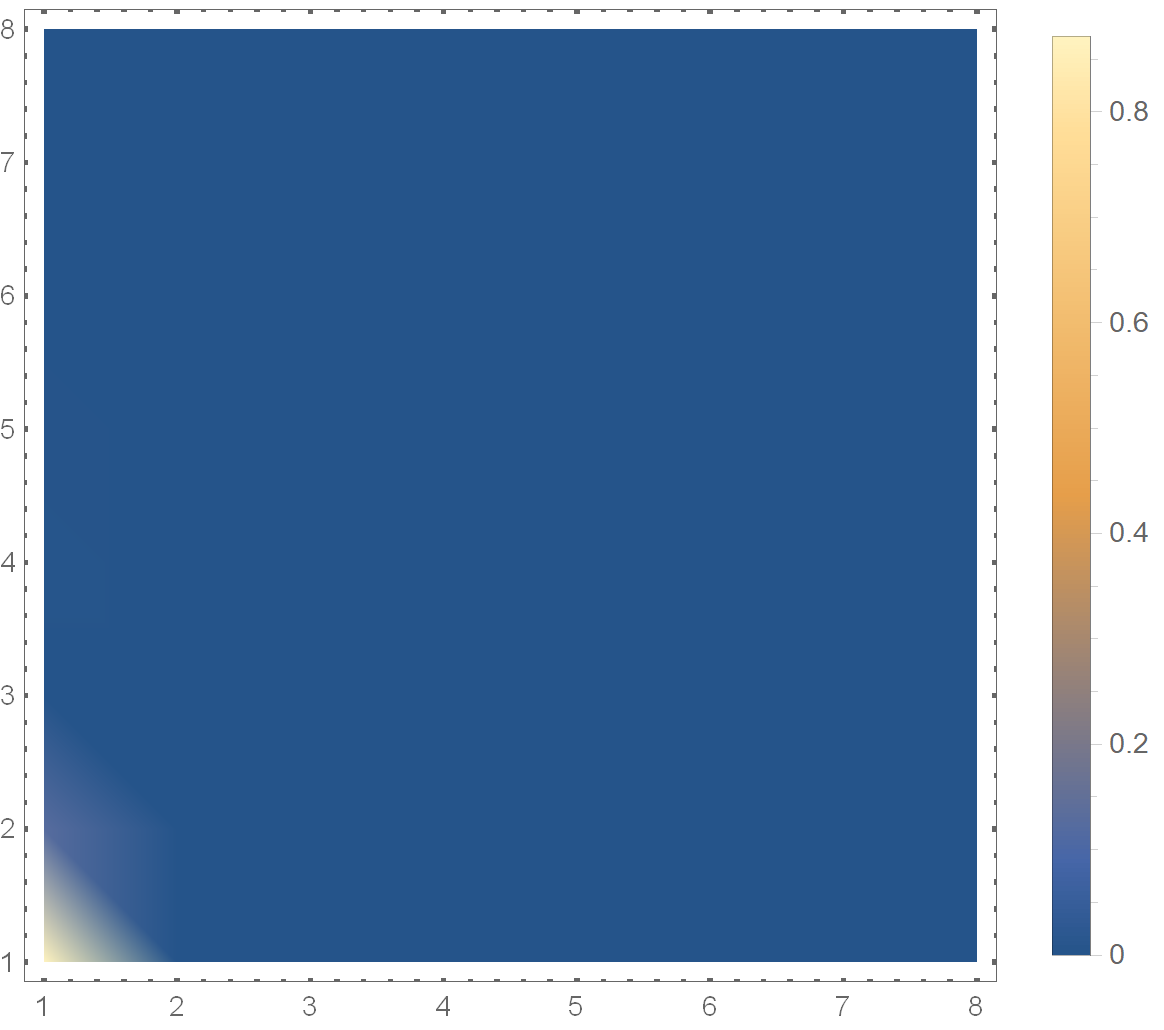} \label{fig2f}}
        \quad
        \subfigure[]{\includegraphics[scale=0.23]{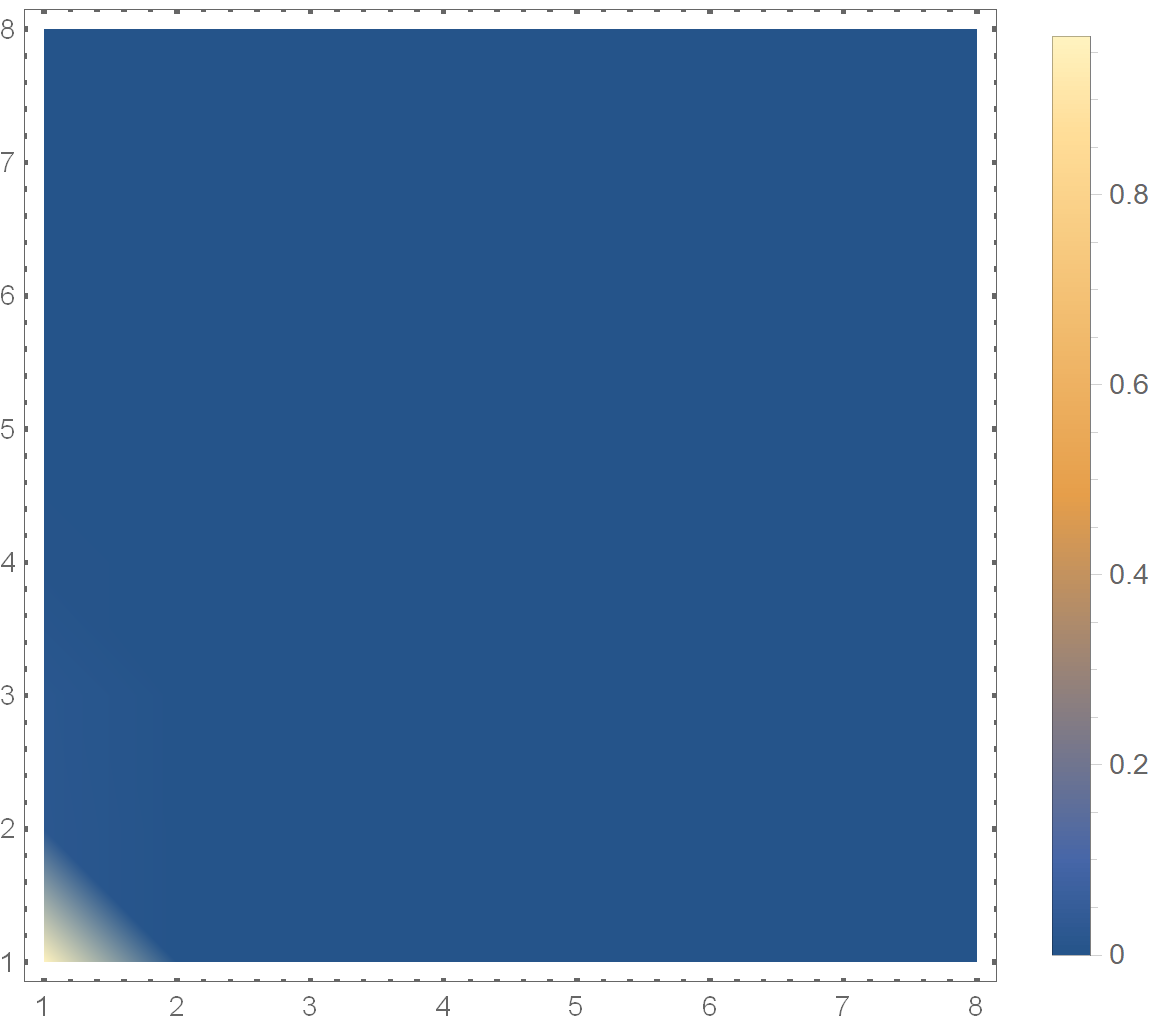} \label{fig2g}}
        \quad
        \subfigure[]{\includegraphics[scale=0.23]{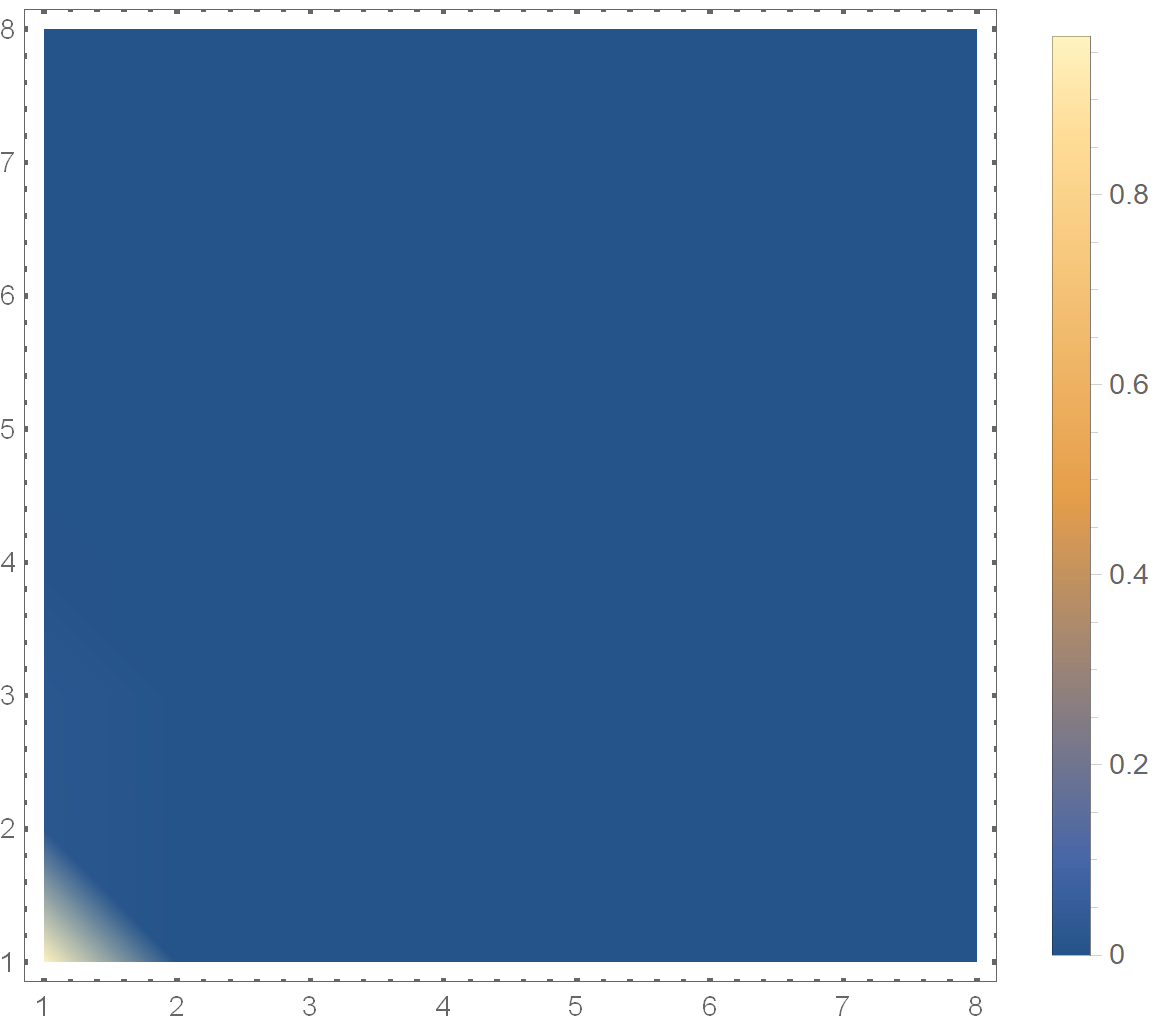} \label{fig2h}}
        \caption{(a)-(h) Density of all $8$ wave functions of the system for $L=8$, $t_x = \gamma_x = 1$, $t_y =2$, $\gamma_y = \frac{3}{2}$, $t_1 = \frac{1}{2}$, $t_2 = \frac{5}{4}$, $t_3 = t_4 = 0$, which are all localized at $x=1$.}
        \label{figs2}
     \end{figure}

\begin{figure}
       \centering
       \subfigure[]{\includegraphics[scale=0.23]{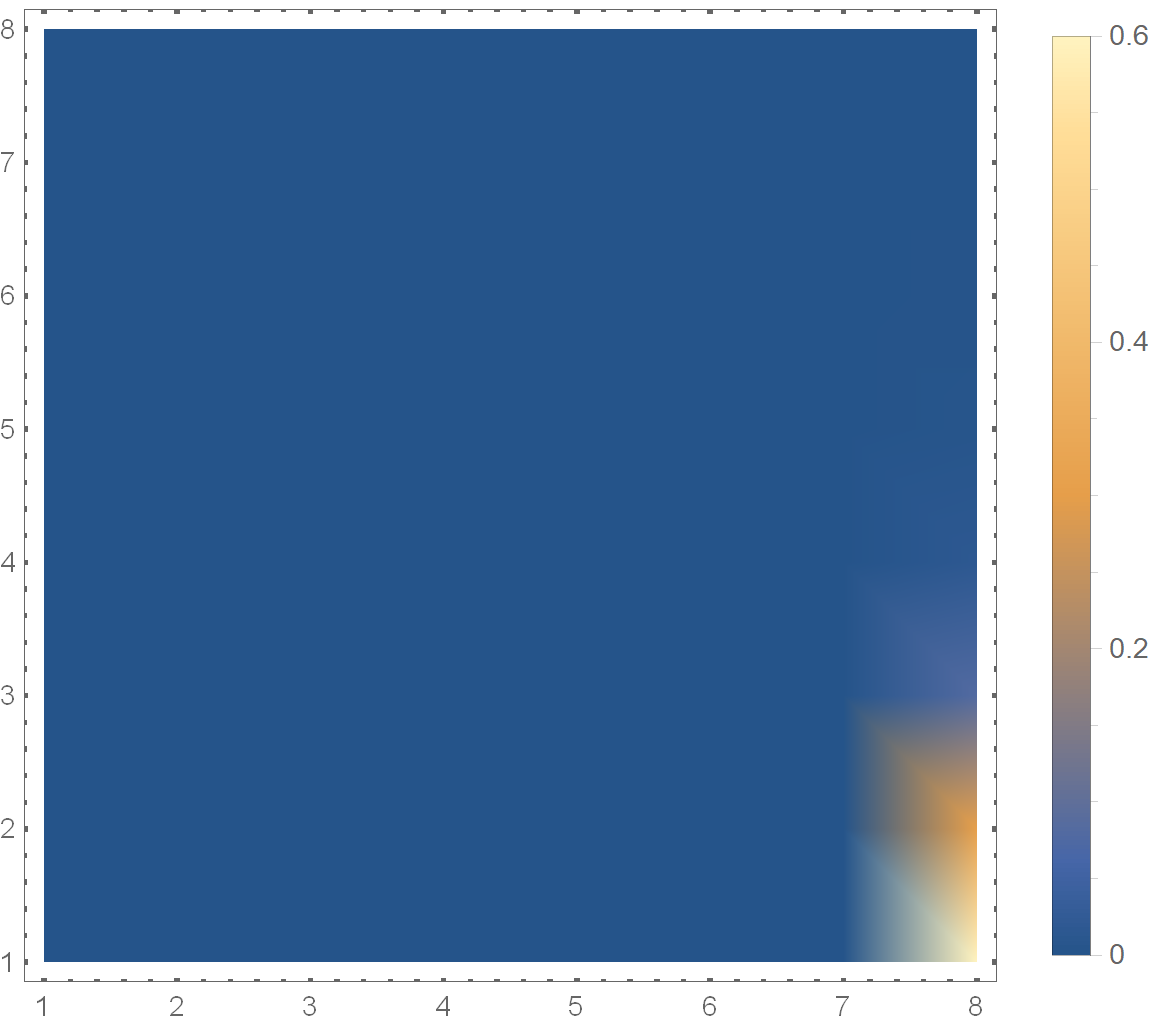} \label{fig3a}}
       \quad
       \subfigure[]{\includegraphics[scale=0.23]{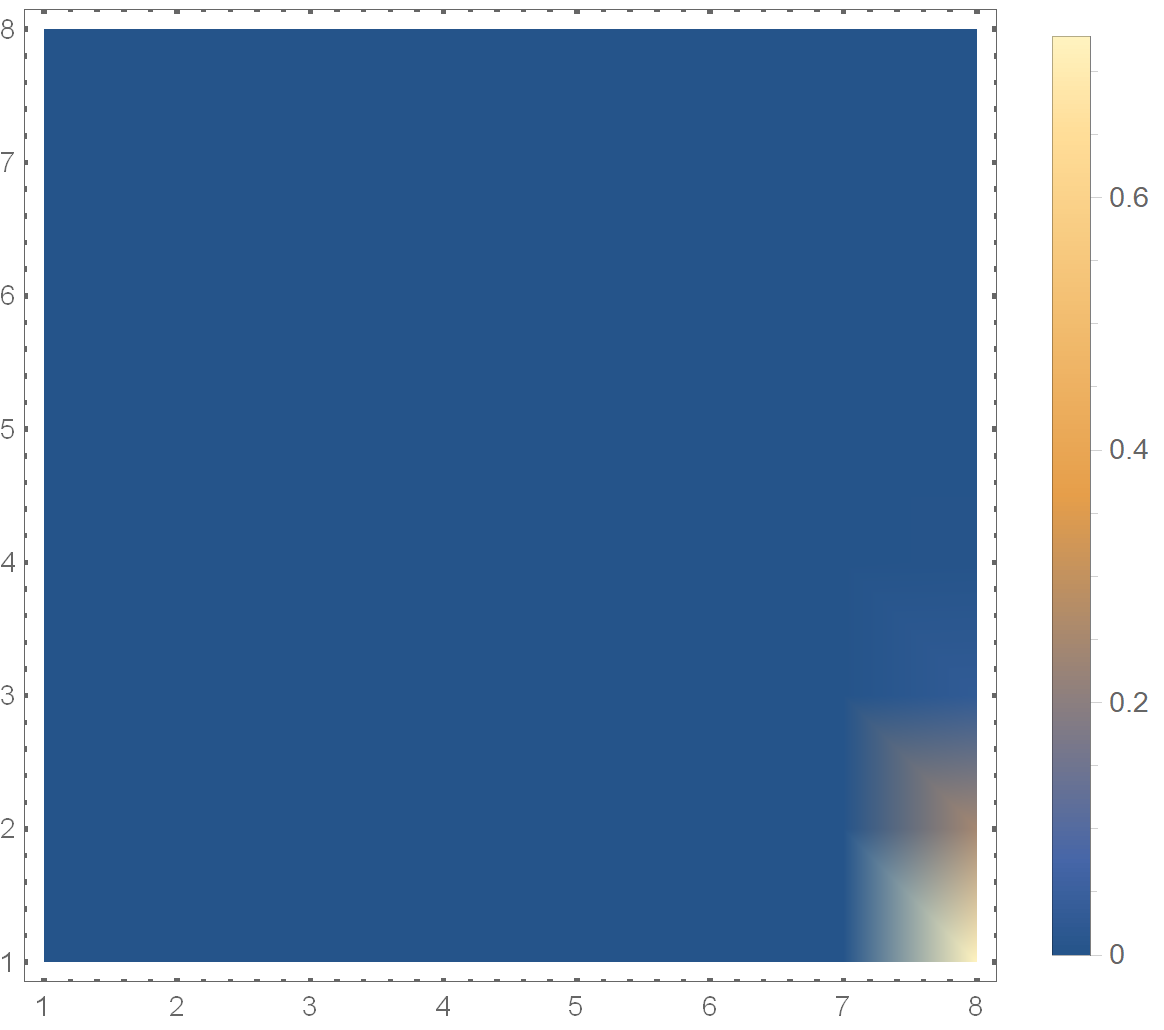} \label{fig3b}}
       \quad
       \subfigure[]{\includegraphics[scale=0.23]{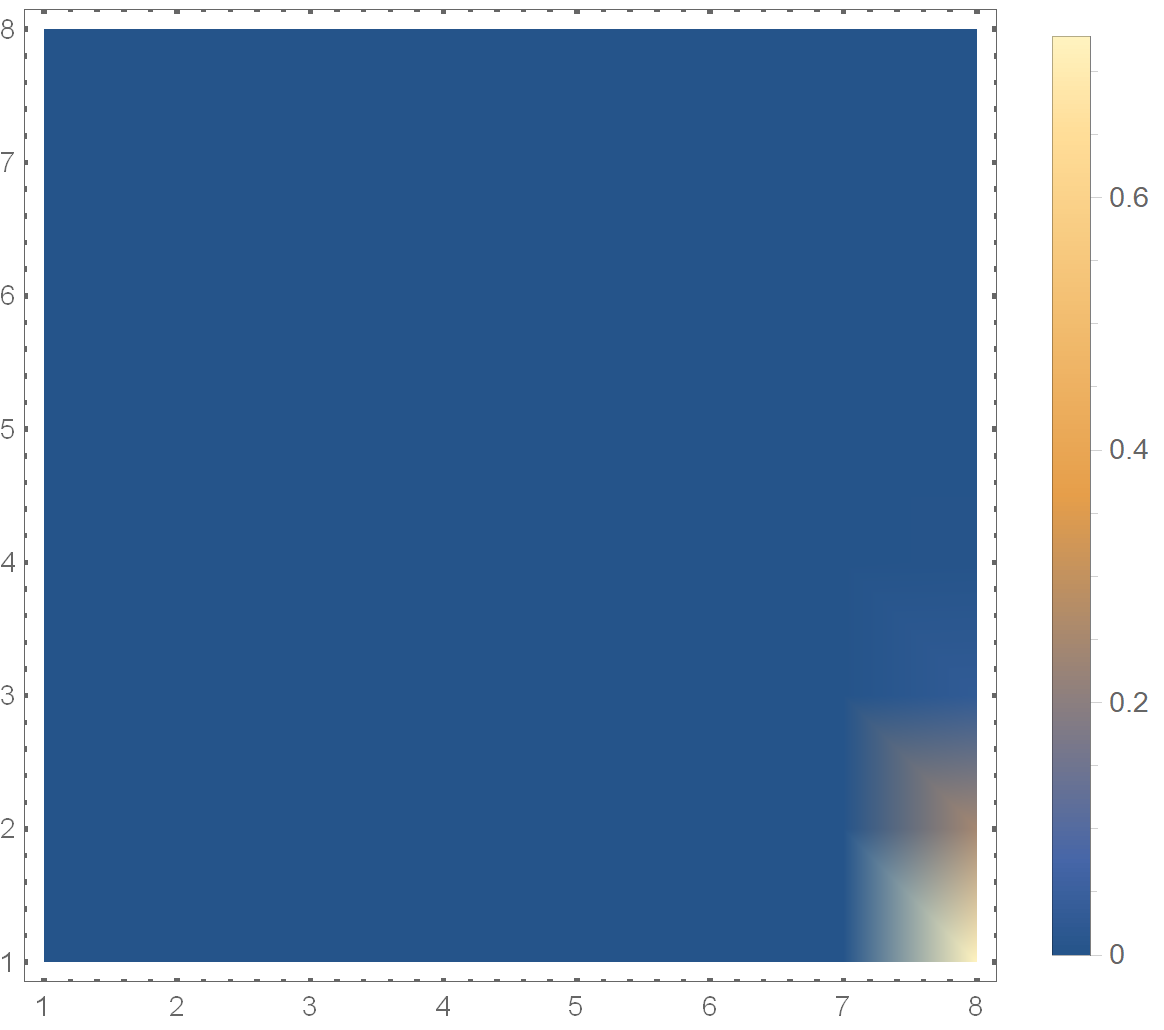} \label{fig3c}}
       \quad
       \subfigure[]{\includegraphics[scale=0.23]{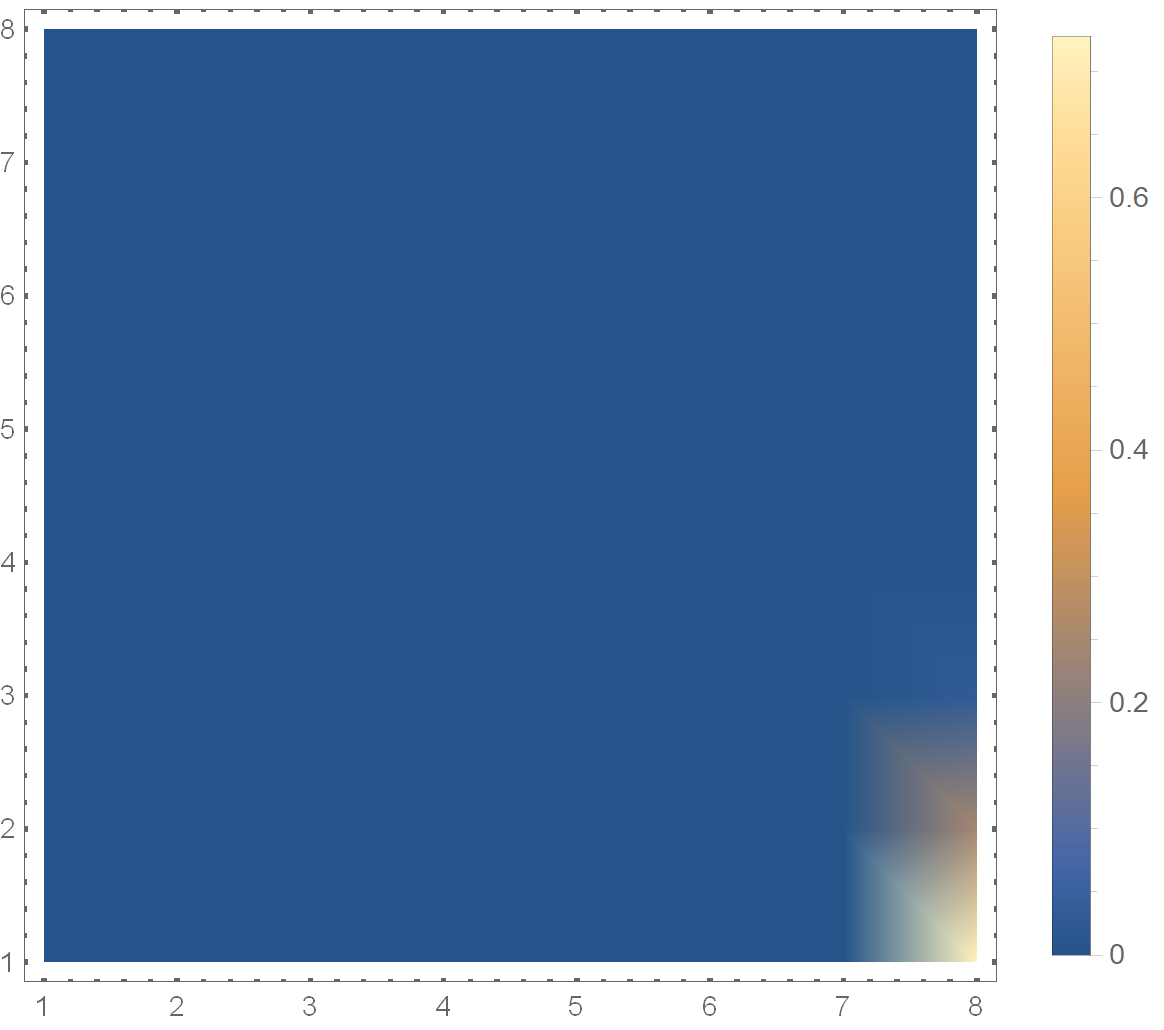} \label{fig3d}}
       \quad
       \subfigure[]{\includegraphics[scale=0.23]{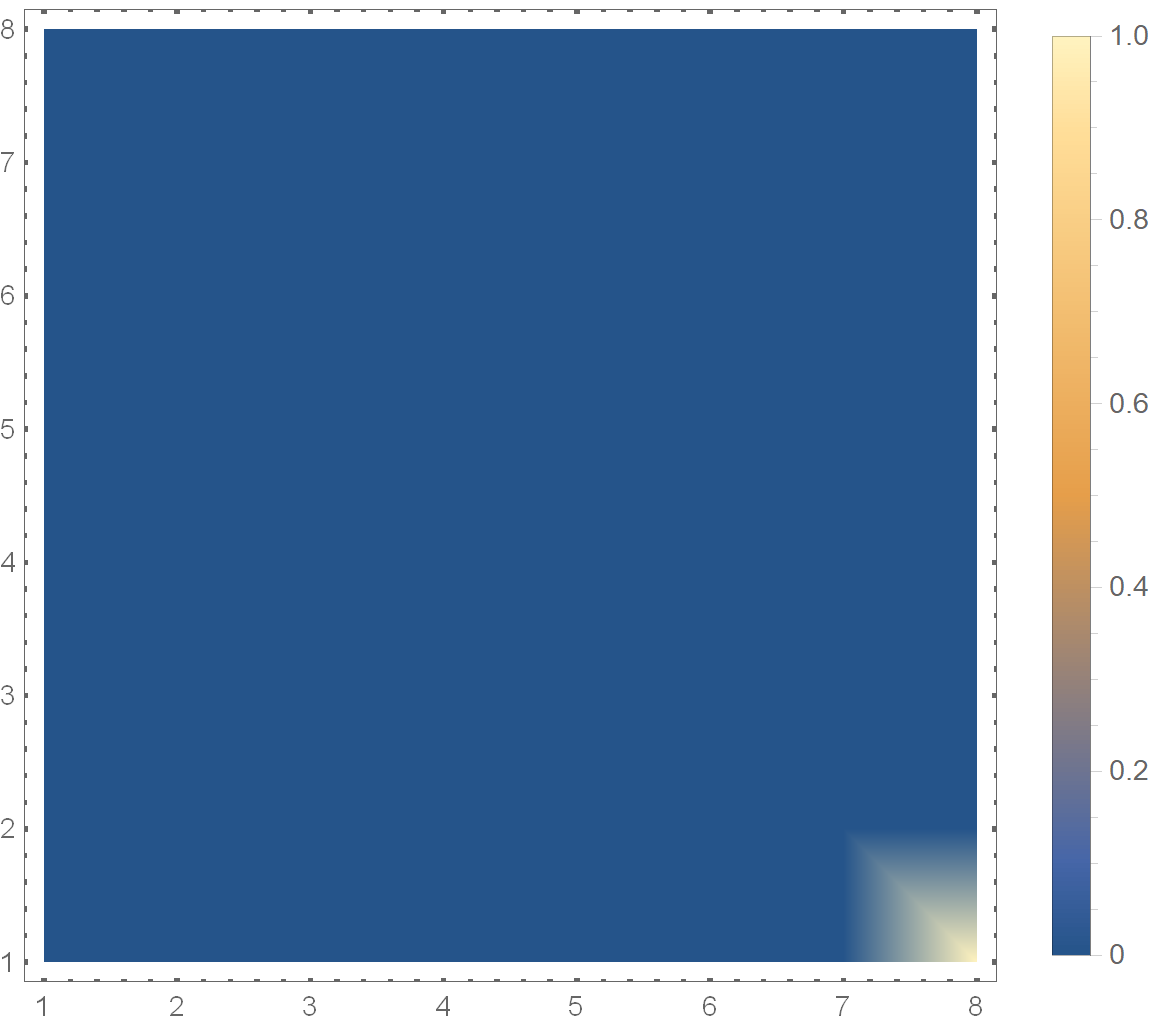} \label{fig3e}}
       \quad
       \subfigure[]{\includegraphics[scale=0.23]{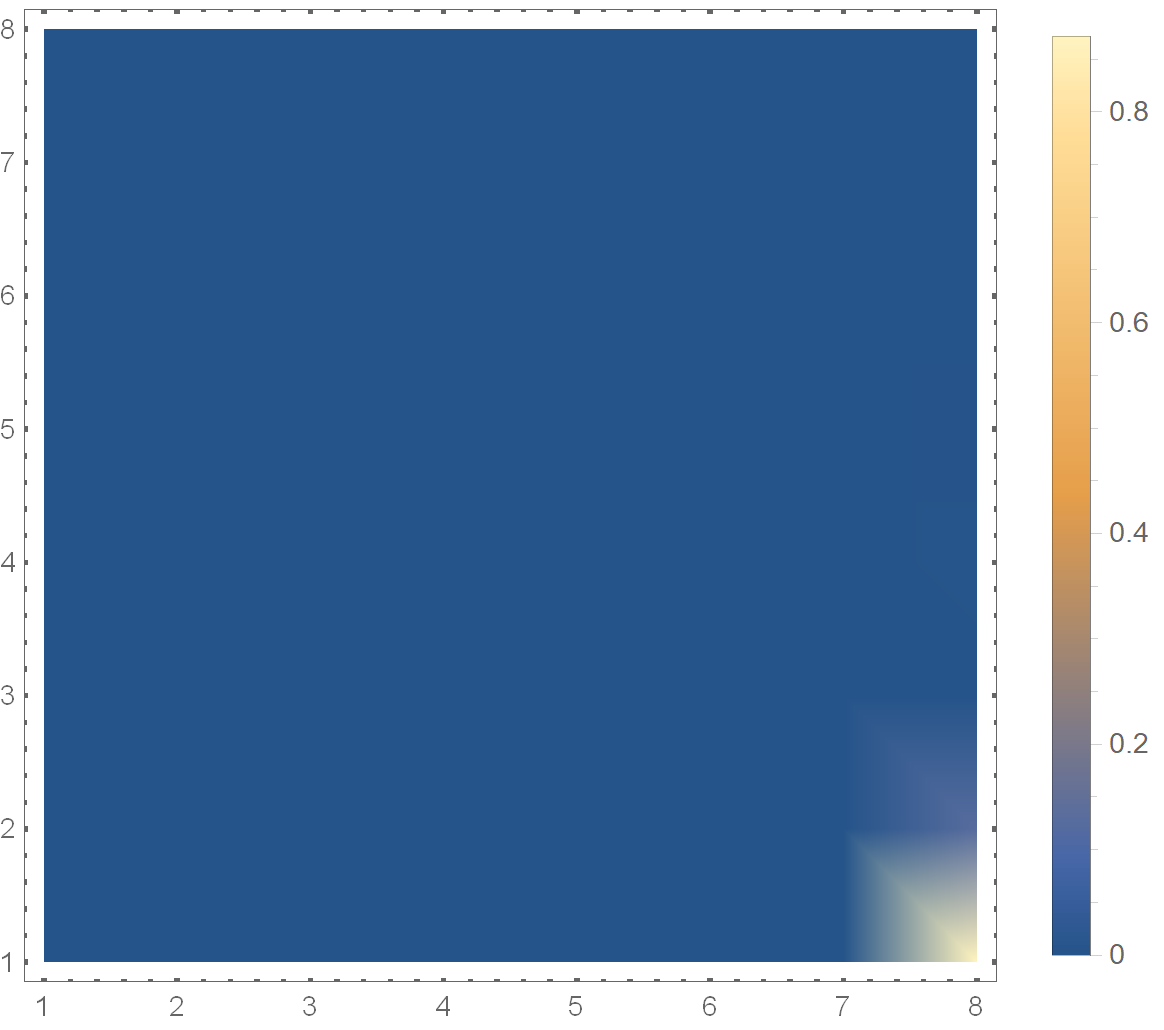} \label{fig3f}}
       \quad
       \subfigure[]{\includegraphics[scale=0.23]{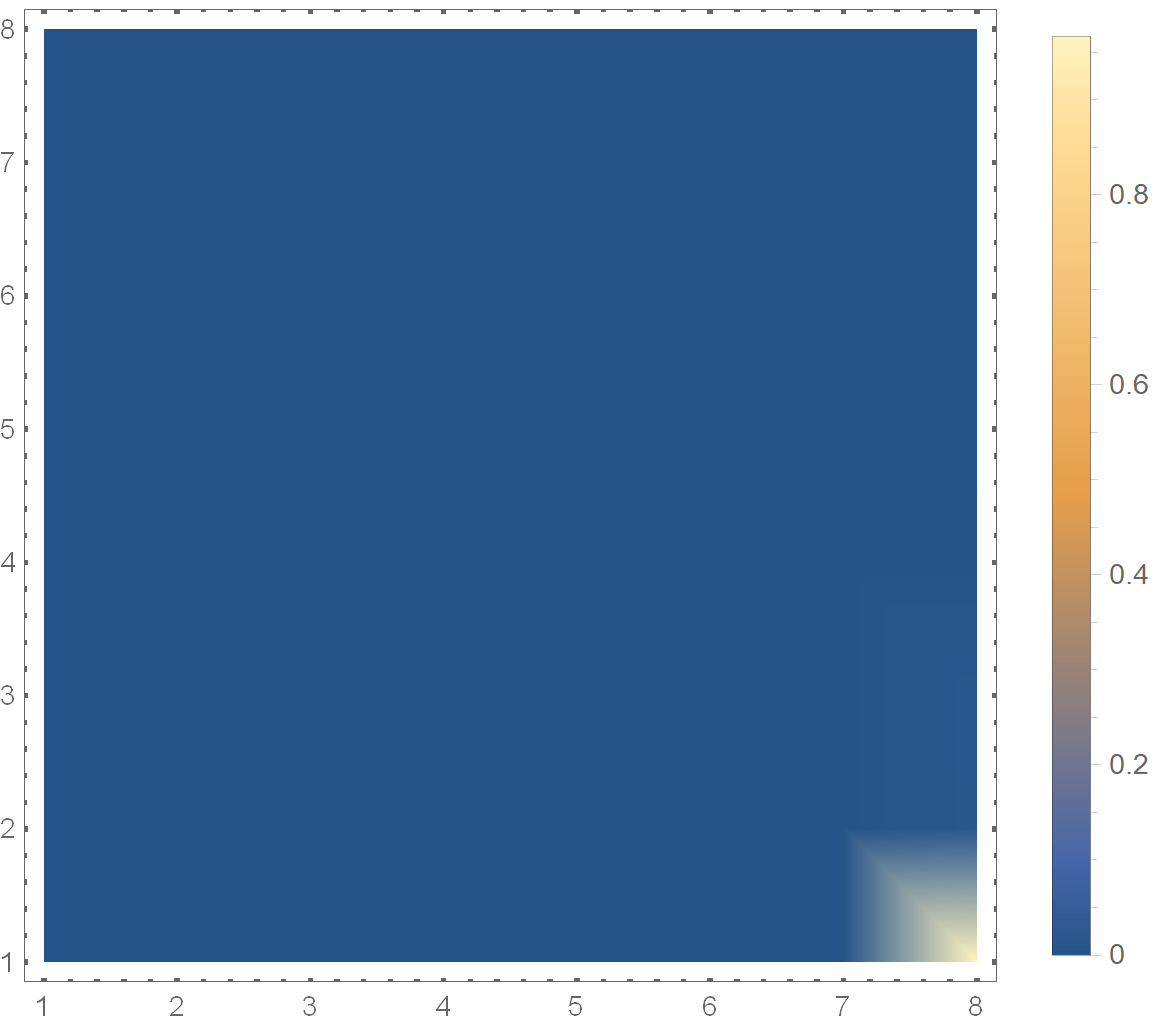} \label{fig3g}}
       \quad
       \subfigure[]{\includegraphics[scale=0.23]{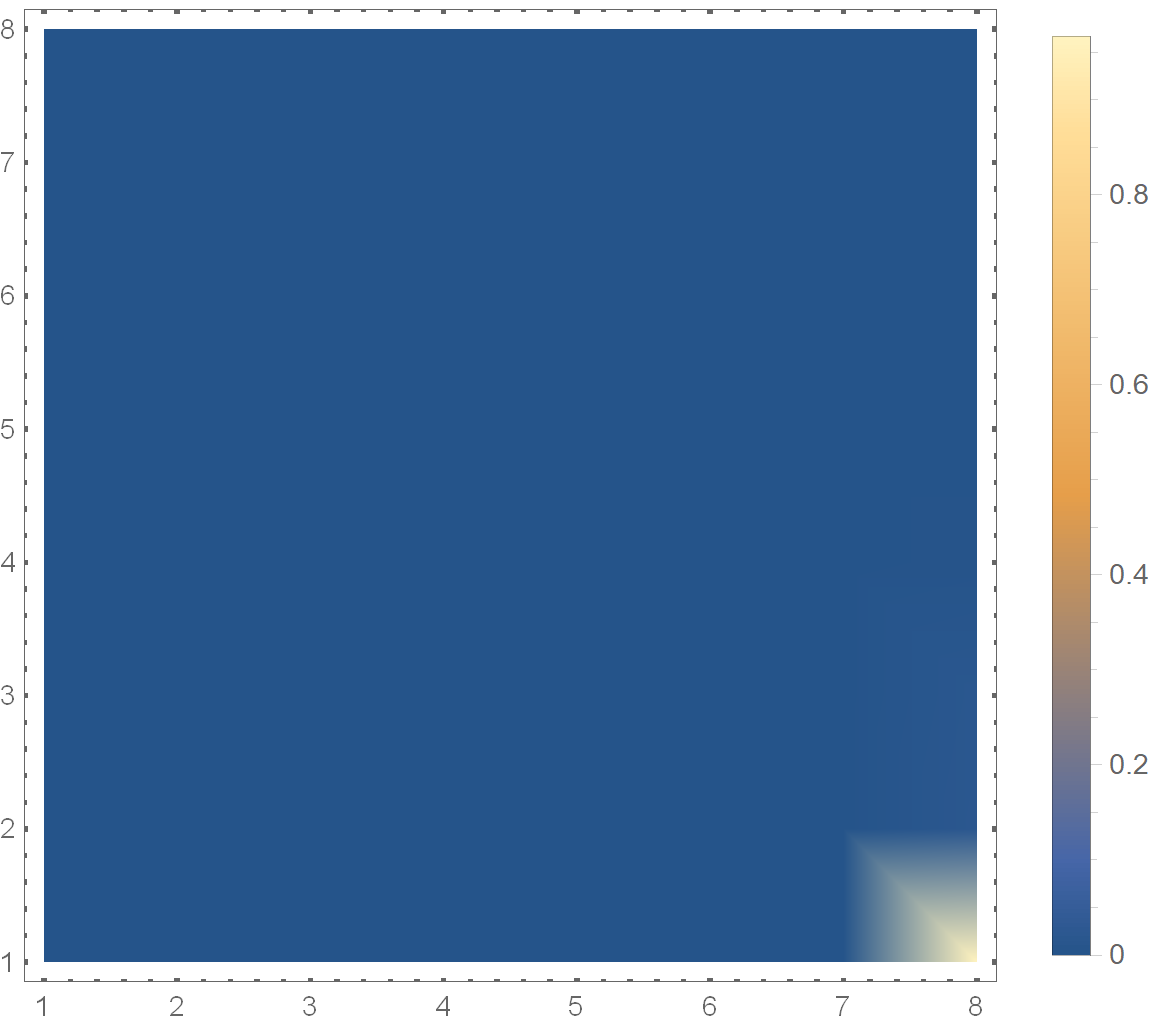} \label{fig3h}}
       \caption{(a)-(h) Density of all $8$ wave functions of the system for $L=8$, $t_x = -\gamma_x = 1$, $t_y =2$, $\gamma_y = \frac{3}{2}$, $t_3 = \frac{1}{2}$, $t_4 = \frac{5}{4}$, $t_1 = t_2 = 0$, which are all localized at $x=8$.}
       \label{figs3}
    \end{figure}

     The characteristic polynomial can be written as
       \begin{equation}
           \begin{split}
             f_2  ( \beta_x) = & \left( t_x - \gamma_x + t_3 \beta_y +t_4 \beta_y^{-1} \right) \beta_x^{-1} + \left( t_x + \gamma_x + t_1 \beta_y + t_2 \beta_y^{-1} \right) \beta_x
             \\
             & + (t_y + \gamma_y) \beta_y + (t_y - \gamma_y) \beta_y^{-1} -E
             \\
             =& \alpha_{x, -1} \beta_{x}^{-1} +\alpha_{x, 1} \beta_{x}^{1} +\alpha_{x, 0} .
           \end{split}
           \tag{S74}
           \label{S74}
       \end{equation}
     Take $L=8$, $t_x = \gamma_x = 1$, $t_y =2$, $\gamma_y = \frac{3}{2}$, $t_1 = \frac{1}{2}$, $t_2 = \frac{5}{4}$, $t_3 = t_4 = 0$. In this case, $\alpha_{x, -1}=0$ and we find that the system has only $8$ eigenstates which are all localized at $x=1$~(Fig.\ref{figs2}).
     Similarly, if we take $L=8$, $t_x = -\gamma_x = 1$, $t_y =2$, $\gamma_y = \frac{3}{2}$, $t_3 = \frac{1}{2}$, $t_4 = \frac{5}{4}$ and $t_1 = t_2 = 0$, $\alpha_{x, 1}=0$. In this case, the system still has only $8$ eigenstates, but all eigenstates are localized at $x=8$~(Fig.\ref{figs3}).
     \par

     According to our theory, infernal points appear if $\alpha_{x, -1}=0$ or $\alpha_{x, 1}=0$ for this system, which is consistent with the example here.

\end{document}